\documentclass[fleqn,usenatbib]{mnras}
\usepackage{newtxtext,newtxmath}
\usepackage[T1]{fontenc}
\usepackage{ae,aecompl,bm,xcolor}
\usepackage{amsmath,color,xspace,multirow,longtable,graphicx,bigints,comment}
\usepackage{tabularx}
\newcommand\setrow[1]{\gdef\rowmac{#1}#1\ignorespaces}
\newcommand\clearrow{\global\let\rowmac\relax}
\clearrow

\newcommand{\cii}{[C\,II]\xspace}
\newcommand{\spt}{SPT\,0346-52\xspace}
\newcommand\textlcsc[1]{\textsc{\MakeLowercase{#1}}}

\title[Gas and Star Formation from HD and Dust...]{Gas and Star Formation from HD and Dust Emission in a Strongly Lensed Galaxy}
\author[G. C. Jones et al.]{
G. C. Jones,$^{1,2}$\thanks{E-mail: gj283@cam.ac.uk}
R. Maiolino,$^{1,2}$,
P. Caselli$^3$,
S. Carniani$^4$
\\
$^{1}$Cavendish Laboratory, University of Cambridge, 19 J. J. Thomson Ave., Cambridge CB3 0HE, UK\\
$^{2}$Kavli Institute for Cosmology, University of Cambridge, Madingley Road, Cambridge CB3 0HA, UK\\
$^3$Centre for Astrochemical Studies, Max-Planck-Institute for Extraterrestrial Physics, Giessenbachstrasse 1, 85748 Garching, Germany\\
$^4$Scuola Normale Superiore, Piazza dei Cavalieri 7, I-56126 Pisa, Italy
}

\date{Accepted 2020 September 01. Received 2020 August 28; in original form 2020 June 25}

\begin{document}
\label{firstpage}
\pagerange{\pageref{firstpage}--\pageref{lastpage}}
\maketitle

\begin{abstract}
The molecular gas content of high-redshift galaxies is a highly sought-after property. However, H$_2$ is not directly observable in most environments, so its mass is probed through other emission lines (e.g., CO, [CI], [CII]), or through a gas-to-dust ratio. Each of these methods depends on several assumptions, and are best used in parallel. In this work, we extend an additional molecular gas tracer to high-redshift studies by observing hydrogen deuteride (HD) emission in the strongly lensed $z=5.656$ galaxy SPT0346-52 with ALMA. 
While no HD(1-0) emission is detected, we are able to place an upper limit on the gas mass of $\rm M_{H_2}<6.4\times10^{11}\,M_{\odot}$. This is used to find a limit on the $L'_{CO}$ conversion factor of $\rm\alpha_{CO}<5.8$\,M$_{\odot}$(K\,km\,s$^{-1}$\,pc$^2$)$^{-1}$. 
In addition, we construct the most complete spectral energy distribution (SED) of this source to date, and fit it with a single-temperature modified blackbody using the nested sampling code MultiNest, yielding a best-fit dust mass $M_{\rm dust}=10^{8.92\pm0.02}$\,M$_{\odot}$, dust temperature $78.6\pm0.5$\,K, dust emissivity spectral index $\beta=1.81\pm0.03$, and star formation rate $SFR=3800\pm100$\,M$_{\odot}$\,year$^{-1}$.
Using the continuum flux densities to estimate the total gas mass of the source, we find M$_{H_2}<2.4\times10^{11}$\,M$_{\odot}$, assuming sub-solar metallicity. This implies a CO conversion factor of $\alpha_{\rm CO}<2.2$, 
which is between the standard values for MW-like galaxies and starbursts. 
These properties confirm that SPT0346-52 is a heavily starbursting, gas rich galaxy.
\end{abstract}

\section{Introduction}
Molecular hydrogen (H$_2$) is both the most abundant molecule in the universe and the main fuel for star formation (e.g., \citealt{kenn12}). However, direct observations of H$_2$ are difficult, as the symmetry inherent in its structure dictates that line emission only originates from energetic environments. To work around this limitation, emission from other molecules may be used as a tracer. 

The main alternative is the second most abundant molecule, carbon monoxide (CO), which is $10^{4}$ times less abundant than H$_2$ (e.g., \citealt{bola13}). The mass of molecular hydrogen in a source may be estimated by observing CO (J=1-0) emission and assuming a conversion factor $\alpha_{CO}=M_{H_2}/L'_{CO}$[=]M$_{\odot}$(K\,km\,s$^{-1}$\,pc$^2$)$^{-1}$ , which is assumed to be $\sim0.8$ for starburst galaxies and $\sim4.6$ for relatively quiescent, Milky Way-like galaxies (e.g., \citealt{dadd10}). But these conversion factors are strongly metallicity dependent (e.g., \citealt{nara12}), and observations of higher-J lines must be converted to the J=1-0 line by the assumption of an uncertain factor \citep{cari13}. Alternatively, one may observe emission from warm dust, convert this luminosity to a dust mass (e.g, \citealt{lamp19}), and convert this to a gas mass by assuming a gas to dust conversion factor ($\delta_{\rm gd}$, e.g. \citealt{sain13}).

A third alternative, which is still largely unexploited, is to observe hydrogen deuteride(HD), which may be converted directly into an H$_2$ mass using the HD/H$_2$ ratio at the observed redshift (e.g., \citealt{berg13}). In regions where the gas is mainly molecular, i.e. when hydrogen is in molecular form, deuterium is mainly locked in HD. HD is rapidly photodissociated in diffuse clouds \citep{spit73} and at the edge of photo-dissociation regions (PDRs), where
the visual extinction falls below $\sim2-3$\,mag \citep{jans95}. However, virtually all of the deuterium is expected to be contained within HD in dense molecular clouds in our Galaxy \citep{tiel05} and beyond, as it can efficiently form both in the gas phase via H$_2$+D$^+$ at low metallicities and on the surface of dust grains even
at dust temperatures above 150\,K (\citealt{wats73,caza09}).

HD has been regularly detected via absorption of electronic
transitions towards stars in the Milky Way (e.g., \citealt{spit73,laco05}) and towards other galaxies (e.g., \citealt{note08,note10,bala10,ivan10,oliv14,dapr17}). Rotational transitions of HD, which can provide mass estimates, have not yet been observed in external galaxies. Only in the Milky Way there have been detections of the HD J=1-0 ground state rotational transition at 112.2\,$\mu$m toward the photodissociation region Orion Bar using the Infrared Space Observatory ISO (e.g. \citealt{wrig99}), and toward protoplanetary disks using the Herschel Space Observatory (\citealt{berg13,mccl16}). \citet{neuf06} also detected the excited HD J=4-3 and 5-4 transitions (both good pressure probes) at 28.5\,$\mu$m and 23.1\,$\mu$m, respectively, toward a supernova remnant using the Spitzer Space Telescope.

\citet{berg13} used HD(1-0) to derive an important lower limit of the TW Hya protoplanetary disk mass, taking into account the fact that some atomic D could be hidden in molecular ices instead of HD; they found significantly larger masses than previously derived with CO (and rare CO isotopologue) lines and with dust continuum emission. It is therefore important to extend the study of HD rotational transitions to external galaxies and compare the new derived values of the gas masses with those obtained with the more classical CO and dust emission methods. High-redshift galaxies allow us to search for the lowest J transitions of HD using sensitive ground based telescopes, such as ALMA. As clearly shown by \citet{berg13}, the bulk molecular gas reservoir of molecular hydrogen can be directly assessed using HD(1-0), once the physical structure  (volume density and kinetic temperature) is known. Unlike CO, HD is expected to have a constant abundance relative to H$_2$ throughout the dense molecular material and emission is expected as long as the gas temperature is above $\sim10-15$\,K. Therefore, with a good knowledge of the (average) density and temperature toward a high-redshift galaxy, HD(1-0) provides a good alternative method to estimate the gas mass in these distant objects. To test this, we search for emission from the fundamental rotational transition HD(J=1-0) from the energetic, strongly lensed galaxy \spt.

SPT-S J034640-5204.9 (hereafter \spt) is a strongly lensed dusty starforming galaxy (DSFG) at $z=5.656$, first studied in the ALMA survey of \citet{weis13} and \citet{viei13}. Detailed gravitational lens modeling shows that the galaxy is magnified by a factor $\mu=5.6\pm0.1$ \citep{spil16}, while a source-plane reconstruction strongly resembles either an ongoing major merger or rotating disk (\citealt{spil15,litk19,dong19}). The detection of a central outflow coincident with an extreme starburst rejects the merger scenario, adding additional credence to the rotator scenario \citep{jone19}. SED modeling yields a massive star formation rate ($SFR\sim4000-5000$\,$M_{\odot}$\,year$^{-1}$, \citealt{ma15}) and further observations have found substantial \cii\,158\,$\mu$m ($L_{[C\,II]}=(5.0\pm0.7)\times10^{10}\,L_{\odot}$, \citealt{gull15}) and CO ($L_{CO(2-1)}=(2.4\pm0.2)\times10^{8}\,L_{\odot}$, \citealt{arav16}) emission. However,  X-ray and radio observations have not revealed the presence of any AGN and suggest that this source is mainly powered by star formation \citep{ma16}.

In this work, we present new ALMA observations in band 8, resulting in a upper limit on the intensity of HD(1-0) emission. In addition, we add our new continuum flux densities to those previously published, and fit the resulting SED with a blackbody submm-FIR model, yielding new estimates on the star formation rate, dust temperature, dust mass, and several limits on the molecular gas mass, resulting in limits for the CO-to-H$_2$ conversion factor. We will assume ($\Omega_{\Lambda}$,$\Omega_m$,h)=(0.692,0.308,0.678) \citep{plan16} throughout.

\section{Observations and Data Reduction}\label{datacal}

Using band 8 of ALMA, we observed \spt from 24 October - 1 November, 2016, using 41-46 antennas. Out of a total observation time of 12.9\,hours, 6.4\,hours were on-source. Our complex gain, bandpass, and flux calibrators were J0253-5441, J0522-3627, and J0519-4546, respectively. The data were originally calibrated by ALMA staff, but excessive flagging of edge channels resulted in a gap in the resulting frequency coverage. To correct this, we recalibrated the data, flagging only 3 channels at each edge of the spectral windows, rather than the default of 10 channels. 

Our frequency range was covered by two sidebands, each composed of two spectral windows (SPWs), made of 128 channels of width 15.625\,MHz. The lower sideband ($387.562-391.031$\,GHz) was expected not to contain prominent emission or absorption lines, while the redshifted frequency of the HD(1-0) transition ($\nu_{rest}=2674.986094$\,GHz, $\nu_{obs}=401.891$\,GHz) falls in the upper sideband ($399.757-403.351$\,GHz).

Since the goal of these observations is to detect HD(1-0) emission, we first create full (i.e., line and continuum) data cubes using the CASA task \textlcsc{tclean}, natural weighting, and a clean threshold of $3\sigma$, where $\sigma$ is the RMS noise level per channel of the dirty image. We explore both the native channel width and 2, 3, 4, and 5-channel averaging. None of these data cubes shows an obvious HD(1-0) signal, so we proceed with the native channel width.

Since the emission line is either weak or nondetected, we perform continuum subtraction, in order to isolate any low-level signal. This process is non-trivial, as the atmospheric transmission across our sidebands varies dramatically\footnote{https://almascience.eso.org/about-alma/atmosphere-model} and the velocity width of the HD(1-0) line is unknown. While the lower (continuum-only) sideband would ideally be used to estimate the continuum level, it shows only $30\%$ transmission (assuming 2.0\,mm PWV), while the upper (HD) sideband has a more favorable transmission ($\sim50\%$), but is plagued by multiple atmospheric absorption lines. 

In order to separate the possible line and continuum emission in these data, we explored multiple continuum subtraction techniques. First, we used the CASA task \textlcsc{uvcontsub} to fit and subtract a first-order polynomial continuum model directly to the visibilities of both sidebands, resulting in purely line emission. Since this task fits the continuum emission directly in the visibilities, it is independent of user-provided imaging parameters. When performing this fitting, we avoided the three frequency ranges of atmospheric absorption in the upper sideband (i.e., $\sim400.0$\,GHz, $\sim401.3$\,GHz, $\sim402.4$\,GHz) and all channels that could include HD line emission. Since the linewidth of HD is unknown, we adopted a conservative line channel range of the expected HD frequency $\pm750$\,km\,s$^{-1}$, based on FWHM$_{\rm CO}=613\pm30$\,km\,s$^{-1}$ \citep{arav16}, resulting in theoretical line channels of $400.888-402.899$\,GHz. Using the CASA task \textlcsc{tclean}, the resulting continuum-subtracted visibilities were then imaged using natural weighting and a clean threshold of $3\sigma$. This resulted in an obvious under-subtraction of the continuum, so this continuum subtraction method was disregarded.

As an alternative to visibility-space continuum subtraction, we also explored the use of image-space subtraction by applying the CASA task \textlcsc{imcontsub} (with both 0$^{\rm th}$- and 1$^{\rm st}$-order polynomial fits) to a full data cube spanning both sidebands. This resulted in a nearly identical data cube as the above \textlcsc{uvcontsub} approach and was not used.

The above tests reveal that the lower sideband is unsuitable for continuum fitting, and that our `conservative' channel exclusion is too rigorous. In order to correct this, we re-examine the \textlcsc{imcontsub} approach (both 0th and 1st-order) for the upper sideband, but only excluding $\pm250$\,km\,s$^{-1}$ on either side of the HD(1-0) line. The 1st order approach returns a flatter spectrum, so we proceed with this continuum-subtracted cube.

Our final continuum subtracted cube has a synthesized beam of $0.19''\times0.17''$ at a position angle of $57^{\circ}$, channels of width 15.625\,MHz ($\sim12$\,km\,s$^{-1}$), and an RMS noise level per channel of 0.25\,mJy\,beam$^{-1}$. 

A continuum image was created by applying \textlcsc{tclean} to all line- and atmospheric feature-free channels of both sidebands (i.e., the `conservative' approach of above), natural weighting, multi-frequency synthesis, and a clean threshold of $3\sigma$, where $\sigma$ is the RMS noise level per channel of the dirty image, resulting in a final RMS noise level in the cleaned image of $0.11$\,mJy\,beam$^{-1}$. 

\section{Results}

\subsection{Continuum}\label{contsec}
\begin{figure}
\centering
\includegraphics[width=\columnwidth]{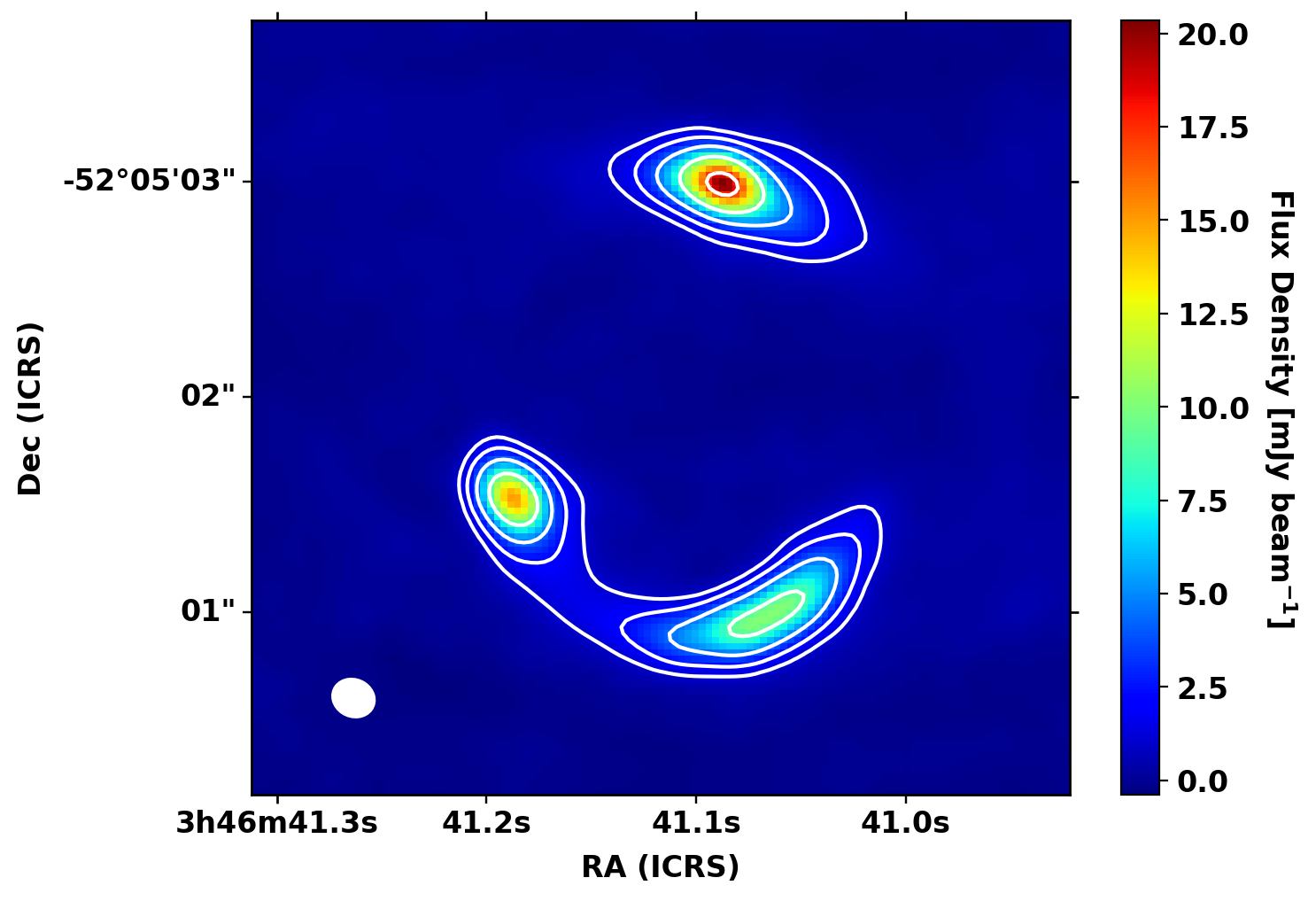}
\caption{The observed continuum ($\rm \lambda_{rest}\sim114\,\mu m$), created using all line-free channels in both sidebands. Contours are shown in a geometric sequence: $[10,20,40,80,160]\times\sigma$, where $\sigma=0.11$\,mJy\,beam$^{-1}$. The synthesized beam ($0.19''\times0.17''$, with major axis position angle = $55^{\circ}$) is shown by the solid white ellipse to the lower left.}
\label{fig:cont}
\end{figure}

The resulting continuum image is shown in Figure \ref{fig:cont}. The continuum is well detected, with a maximum significance of $\sim185\sigma$. The total continuum flux density is $171.5\pm1.0$\,mJy. Note that in this section, we will state the observed (i.e., image-plane) continuum parameters, rather than the de-lensed (or source-plane) values.

A two-dimensional Gaussian fit to the northern component in the continuum image returns a deconvolved size of $(0.39\pm0.02)''\times(0.17\pm0.01)''$ at $(72\pm2)^{\circ}$, an integrated flux density of $56.3\pm2.7$\,mJy, and a peak flux density of $17.9\pm0.7$\,mJy\,beam$^{-1}$.
A similar fit to the southeastern component returns a deconvolved size of $(0.28\pm0.02)''\times(0.16\pm0.02)''$ at $(27\pm6)^{\circ}$, an integrated flux density of $33.6\pm1.9$\,mJy, and a peak flux density of $14.0\pm0.6$\,mJy\,beam$^{-1}$.
Although the southwestern component is extended in an arc, a Gaussian fit returns a deconvolved size of $(0.68\pm0.05)''\times(0.19\pm0.02)''$ at $(117\pm1)^{\circ}$, an integrated flux density of $56.7\pm3.5$\,mJy, and a peak flux density of $9.9\pm0.5$\,mJy\,beam$^{-1}$. Due to lensing effects, these fits are not trivially translatable to physical parameters (e.g., size, surface brightness). However, since the integrated flux density of each fit is higher than its peak, and the spatial scales have small errors, we may state that these sources are well resolved.

Using the $2\sigma$ contour as a spatial mask, we determine the total $\lambda_{\rm rest}\sim114\,\mu$m continuum flux density of this source to be $171.5\pm1.0$\,mJy. This value is greater than the sum of each integrated flux density ($147\pm5$\,mJy), suggesting that the diffuse emission between the components (especially in the south) is significant, or that these resolved components are poorly described by Gaussian fits.

\subsection{HD(J=1-0) Emission} \label{linesec}

Since \spt is strongly detected in multiple gas tracers (i.e.; CO(2-1), \citealt{arav16}; FIR continuum emission), it is expected to contain a substantial amount of molecular gas ($\sim10^{11}$\,M$_{\odot}$), and should therefore be observable in HD(1-0) emission (e.g., \citealt{berg13}). To explore this possibility, we examine a continuum subtracted cube (see Section \ref{datacal} for details of cube creation), searching for significant line emission at the expected redshift. However, the systemic velocity, spatial position, velocity width, and spatial extent of HD(1-0) are not known \textit{a priori}, making this search non-trivial. As a further complication, multiple atmospheric absorption features are present in the observed frequency range (see Section \ref{datacal}), resulting in a sub-optimal continuum subtraction and a varying RMS noise level. By searching the cube, two tentative line features are detected, but they are believed to be noise. For details of these tentative detections ('Tentative 1' and 'Tentative 2'), see Appendix \ref{TD}.

This exploration of the data cube yielded no believable detections, so we turn to the possibility that the HD(1-0) emission is spread over many channels (i.e., broad line width) with a low amplitude. Indeed, previous observations of line emission in \spt found FWHM values of $\sim500-700$\,km\,s$^{-1}$ (e.g., \citealt{arav16,apos19,dong19}), which correspond to $\sim40-60$\,channels in our data cube. If a weak emission feature is distributed over many channels, it is entirely possible that it would not be detected in a channel-by-channel inspection. To test this, we use the CASA task \textit{immoments} to collapse the channels corresponding to $\rm -350<v<350$\,km\,s$^{-1}$ and search for significant features. This collapsed image shows a $3\sigma$ feature that is coincident with the northern FIR continuum image of \spt, but this feature is comparable to noise peaks in the field of view, and is thus likely not significant ('Tentative 3', Appendix \ref{TD}).

\begin{figure}
\centering
\includegraphics[width=\columnwidth]{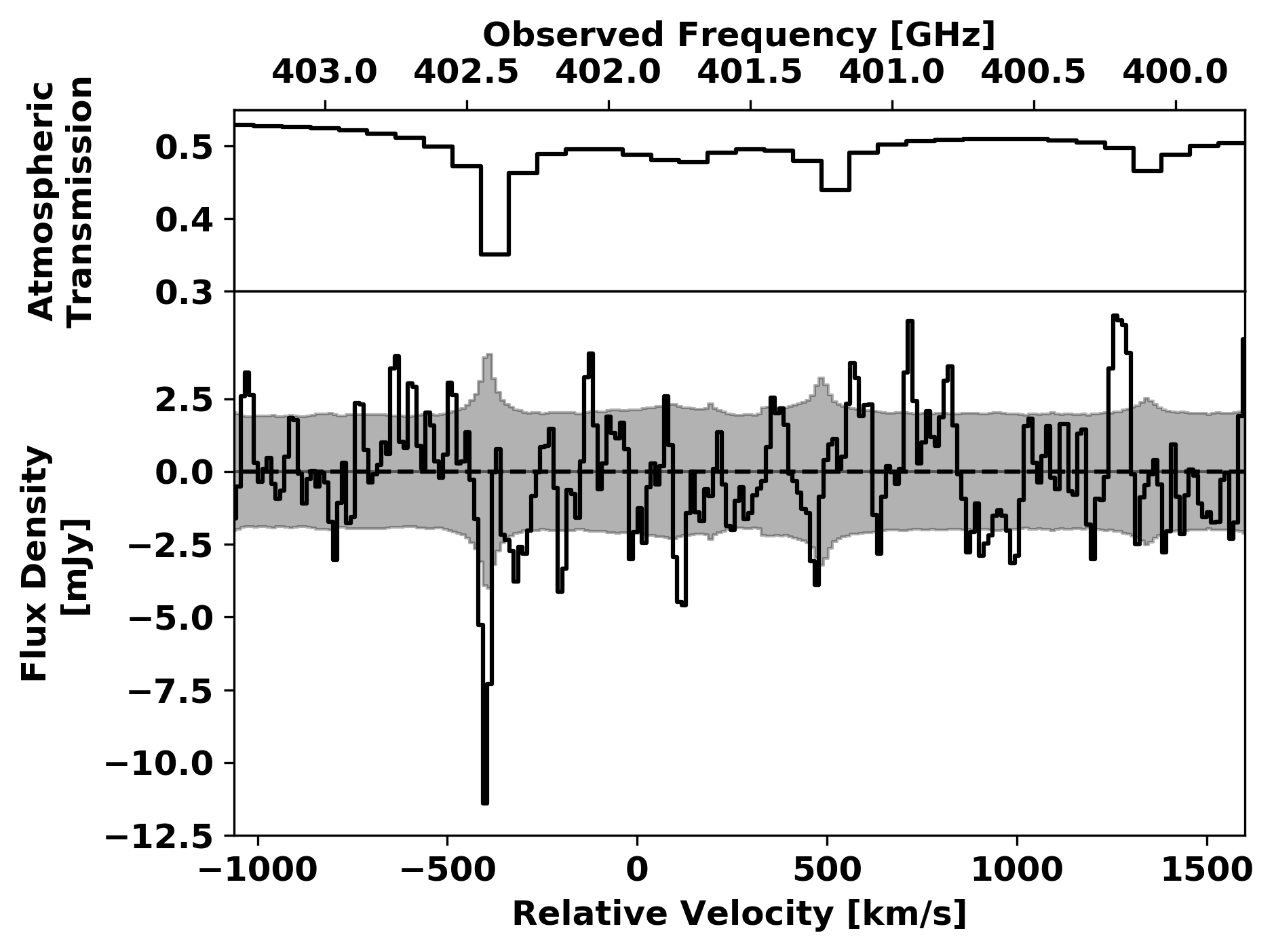}
\caption{Atmospheric transmission curve for 2.0\,mm PWV (top) and spectrum taken over the $2\sigma$ continuum contour (bottom). Grey shaded area in lower panel shows root mean square noise level. Velocity scale is shown relative to the expected frequency of HD(1-0). No lensing correction has been applied.}
\label{fig:rmsatm01}
\end{figure}

Instead, it is plausible that the underlying emission is very spatially extended, on a similar scale as the FIR continuum emission. As shown in Figure \ref{fig:cont}, the FIR continuum is well resolved by our observations, so the emission is spread over multiple observing beams. If HD(1-0) is weak and as extended as the continuum emission, then it is possible that it may only be detected by integrating the emission from a wide area. To test this, we extract a spectrum from the continuum subtracted cube using the $2\sigma$ contour of the continuum map as a mask, resulting in the spectrum shown in Figure \ref{fig:rmsatm01}. A weak feature is detected, but it is at high-velocity, and is thus unlikely to be real ('Tentative 4', Appendix \ref{TD}). 

Since no definite emission is detected through inspecting the data cube, collapsing a wide channel range, or extracting a spectrum from a large area, we conclude that HD(1-0) is undetected in these observations. These observations may be used to place a conservative upper limit on the HD(1-0) flux from \spt. Assuming the same spatial distribution as the FIR continuum emission, we extract a spectrum from our continuum-subtracted cube (Figure \ref{fig:rmsatm01}), finding an RMS noise level of $\sim2.15$\,mJy at the expected frequency. Next, we assume a wide Gaussian line profile (FWHM$_{\rm HD}=613$\,km\,s$^{-1}\sim0.8$\,GHz; \citealt{arav16}) with an amplitude limit of $<2\sigma$, resulting in a conservative upper limit on the integrated intensity of HD(1-0) of $F_L<3.8\times10^{-20}$\,W\,m$^{-2}$. The implications of this nondetection will be discussed in Section \ref{MHM}.

\section{Analysis}

\subsection{SED Modeling} \label{SEDM}

Since \spt is one of the highest-redshift sources detected in the SPT-S survey, and exhibits one of the lowest magnifications in the SPT sample, its intrinsic luminosity is substantial. Due to its extreme characteristics, it has been observed in continuum emission by a host of instruments (e.g., HST, ACT, SPIRE). We compile these observations in Table \ref{tab:sed}.

In an effort to fill in the $\lambda_{\rm obs}\sim1\,$mm regime, we have also compiled multiple continuum detections from the ALMA data archive. For each of these measurements, we use the ALMA-staff calibrated continuum images (i.e., no further calibration has been performed). These continuum measurements may be combined to create a spectral energy distribution (SED), which can be fit with a dust model. 

\subsubsection{Model Description}\label{MD}
We choose a modified blackbody (also known as a greybody; e.g., \citealt{grev12}):
\begin{equation}\label{g12}
S_{obs}(\nu_{obs})=\frac{\Omega}{(1+z)^3}B(\nu,T_{dust})(1-e^{-\tau_{\nu}})
\end{equation}
where $S_{obs}(\nu_{obs})$ is the observed flux density at $\nu_{obs}=\nu_{rest}/(1+z)$, $\Omega=(1+z)^4A_{gal}/D_L^2$ is the solid angle of the galaxy with area $A_{gal}$ at luminosity distance $D_L$, $B(\nu,T_{dust})$ is the blackbody function with a dust temperature $T_{dust}$, and $\tau_{\nu}$ is the optical depth:
\begin{equation}\label{optth}
\tau_{\nu}=\Sigma_{dust}\kappa_{\nu}=\frac{M_{dust}}{A_{gal}}\kappa_o(\nu/\nu_o)^{\beta}
\end{equation}
where we assume the dust absorption coefficient $\kappa_o=0.04$\,m$^2$\,kg$^{-1}$ at $\nu_o=250$\,GHz \citep{beel06}. It is often assumed that the emission is optically thin ($\tau_{\nu}<<1$), which allows equation \ref{g12} to be greatly reduced (e.g., \citealt{case12}). However, we find that this assumption is not applicable for our data (see Section \ref{modres}), and proceed with the general form.

Due to the high redshift of this source, we include the effects of the cosmic microwave background (CMB) on the observed dust continuum, as outlined in \citet{dacu13} and implemented in \citet{carn19}. First, the higher temperature CMB will heat the dust, resulting in a correction on our derived dust temperature:
\begin{equation}
    T_{dust}'=\left[T_{dust}^{4+\beta}+T_o^{4+\beta}\left((1+z)^{4+\beta}-1\right)\right]^{1/(4+\beta)}
\end{equation}
where $T_{dust}$ is the true dust temperature, $T_{dust}'$ is an effective dust temperature, $T_o=2.73$\,K is the CMB temperature at $z=0$, and $\beta$ is the dust emissivity spectral index. In addition to this effect, the hotter CMB provides a background against which we observe the dust emission. This contribution must be removed, resulting in a modified blackbody function:
\begin{multline}
B'(\nu,T_{dust})=B(\nu,T_{dust}')-B(\nu,T_{CMB})\\
= \frac{2h\nu^3}{c^2}\left[\frac{1}{e^{h\nu/k_BT_{dust}'}-1}-\frac{1}{e^{h\nu/k_BT_{CMB}}-1} \right]
\end{multline}
where $T_{CMB}=(1+z)T_o$. 

Combining these equations, we find the following equation for the observed flux density:
\begin{equation}\label{sedsc}
S_{obs}(\nu_{obs})=\frac{(1+z)\pi R_{gal}^2}{D_L^2}B'(\nu,T_{dust})\left( 1-e^{\frac{-M_{dust}\kappa_o(\nu/\nu_o)^{\beta}}{\pi R_{gal}^2}} \right)
\end{equation}
To approximate the radius of emission, we average the source-plane $\lambda_{\rm obs}=$2.0\,mm and 3.0\,mm effective radii of \citet{apos19}, resulting in $R_{gal}\sim0.76$\,kpc.  

Using equation \ref{sedsc}, it is possible to fit the full FIR dust SED with only three free parameters: the dust emissivity spectral index ($\beta$), dust temperature ($T_{dust}$), and dust mass ($M_{dust}$). 

\subsubsection{Model Implementation}
Models were fit to the SED using the Bayesian inference code MultiNest \citep{fero08} and its python wrapper (PyMultiNest; \citealt{buch14}). This code returns the most likely parameter values, parameter value probability distributions, and covariance distributions for each parameter pair. We assume uniform priors for each variable, with log$_{10}$(M$_{\rm dust}/$M$_{\odot})=8-10$, $\beta=1-3$, and T$_{\rm dust}=10-200$\,K. 

However, we must first consider what subset of our sample we may fit with this model. Since we are not including contributions from synchrotron or thermal free-free emission (e.g., \citealt{yun02}), we do not include the low-frequency $\lambda_{\rm obs}=$8\,mm data point.

On the high-frequency side of the model, we choose to include all of the Herschel data ($\lambda_{rest}=24-75$\,$\mu$m), which trace the peak of the dust emission. Previous implementations of a similar model to fit dust SEDs of HyLIRGs (\citealt{yun02,wagg14}) used data from beyond the thermal dust peak. In addition, the SEDs of \citet{ma16} and \citet{apos19} included these points, which were well fit by dust-only models.

Each observed flux density was corrected for magnification by assuming a constant lensing magnification of $\mu=5.6$, based on the detailed modelling of $\lambda_{obs}=870\,\mu$m continuum emission \citep{spil16}.

\begin{table}
\centering
\begin{tabular}{>{\rowmac}c|>{\rowmac}c|>{\rowmac}c|>{\rowmac}c|>{\rowmac}c<{\clearrow}}
$\lambda_{obs}$[$\mu$m] & $\lambda_{rest}$[$\mu$m] & $S_{\nu}$ [mJy] & Instrument & Reference\\ \hline
1.1 & 0.17 & $<3.8\times10^{-4}$ & HST/WFC3 & 1\\
1.6 & 0.24 & $<9.1\times10^{-4}$ & HST/WFC3 & 1\\
3.6 & 0.54 & $<2.4\times10^{-3}$ & Spitzer/IRAC & 1\\
4.5 & 0.68 & $<3.6\times10^{-3}$ & Spitzer/IRAC & 1\\
100 & 15 & $<6.0$ & Herschel/PACS & 1\\
\setrow{\bfseries}160 & 24 & $\bm{33\pm9}$ & Herschel/PACS & 1\\
\setrow{\bfseries}250 & 38 & $\bm{122\pm11}$ & Herschel/SPIRE & 1\\
\setrow{\bfseries}350 & 53 & $\bm{181\pm14}$ & Herschel/SPIRE & 1\\
\setrow{\bfseries}500 & 75 & $\bm{204\pm15}$ & Herschel/SPIRE & 1\\
\setrow{\bfseries}758 & 114 & $\bm{171.5\pm1.0}$ & ALMA & 2\\
\setrow{\bfseries}823 & 124 & $\bm{123.9\pm1.0}$ & ALMA & 3\\
\setrow{\bfseries}870 & 131 & $\bm{123.0\pm13.0}$ & ALMA & 4\\
\setrow{\bfseries}894 & 134 & $\bm{82.3\pm1.7}$ & ALMA & 5\\
\setrow{\bfseries}1028 & 155 & $\bm{79.7\pm1.2}$ & ALMA & 6\\
\setrow{\bfseries}1320 & 198 & $\bm{43.9\pm0.4}$ & ALMA & 7\\
\setrow{\bfseries}1375 & 207 & $\bm{47.1\pm4.3}$ & ACT & 8\\ 
\setrow{\bfseries}1400 & 210 & $\bm{46.0\pm6.8}$ & SPT & 1\\
\setrow{\bfseries}2000 & 301 & $\bm{8.80\pm1.35}$ & ALMA & 9\\
\setrow{\bfseries}2026 & 304 & $\bm{16.7\pm2.6}$ & ACT & 8\\
\setrow{\bfseries}2064 & 310 & $\bm{14.75\pm0.03}$ & ALMA & 10\\
\setrow{\bfseries}3000 & 451 & $\bm{3.06\pm0.05}$ & ALMA & 9\\
8081 & 1214 & $0.16\pm0.02$ & ATCA & 11\\
54508 & 8189 & $<0.114$ & ATCA & 12\\
142758 & 21448 & $<0.213$ & ATCA & 12\\
\end{tabular}
\caption{Continuum flux densities measured for \spt, with no magnification correction. Includes previously published values, values from the ALMA archive, and new values from this paper. Bold entries were used in our SED model.
1- \citet{ma15},
2- This work,
3- 2015.1.01580.S,
4- \citealt{spil16},
5- \citealt{jone19},
6- 2013.1.01231.S,
7- 2016.1.00654.S,
8- \citealt{mars14},
9- \citealt{apos19},
10- \citealt{dong19},
11- \citealt{arav16},
12- \citealt{ma16}.}
\label{tab:sed}
\end{table}

\subsubsection{Model Results}\label{modres}

\begin{figure*}
\centering
\includegraphics[width=\textwidth]{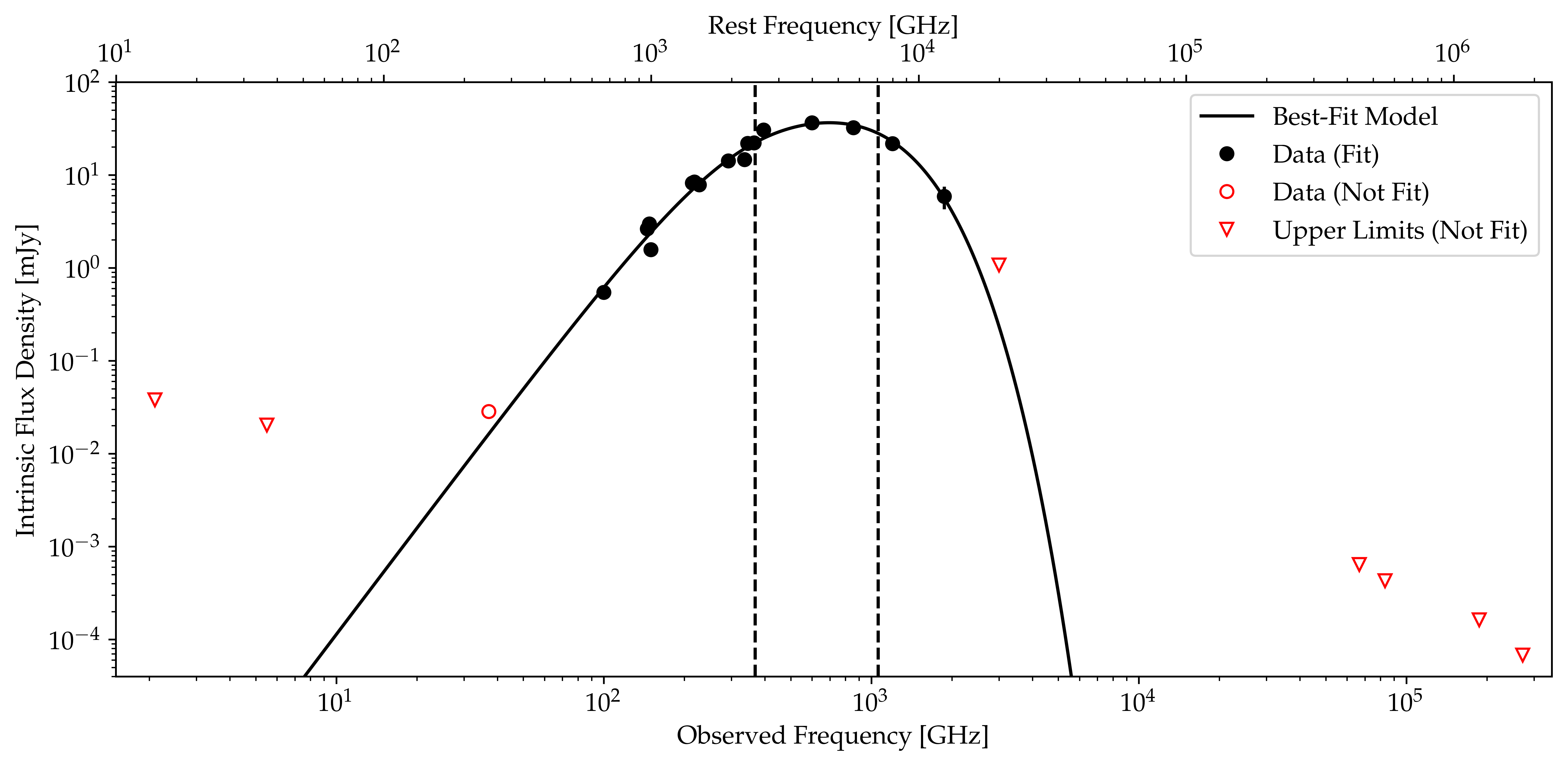}
\caption{The submm-FIR SED for SPT0346-52, showing points from Table \ref{tab:sed}. Data used in the fit are shown in black, while the one excluded low-frequency point and upper limits from literature (which are not used in the fit) are shown by open red circle and triangles, respectively. The flux densities were corrected for lensing magnification by assuming $\mu=5.6\pm0.1$ \citep{spil16}. The best fit modified blackbody model to the sample is shown in black. The frequency range used to measure $L_{FIR}$ is denoted by the vertical black dashed lines.}
\label{fig:sedfit}
\end{figure*}

\begin{figure}
\centering
\includegraphics[width=\columnwidth]{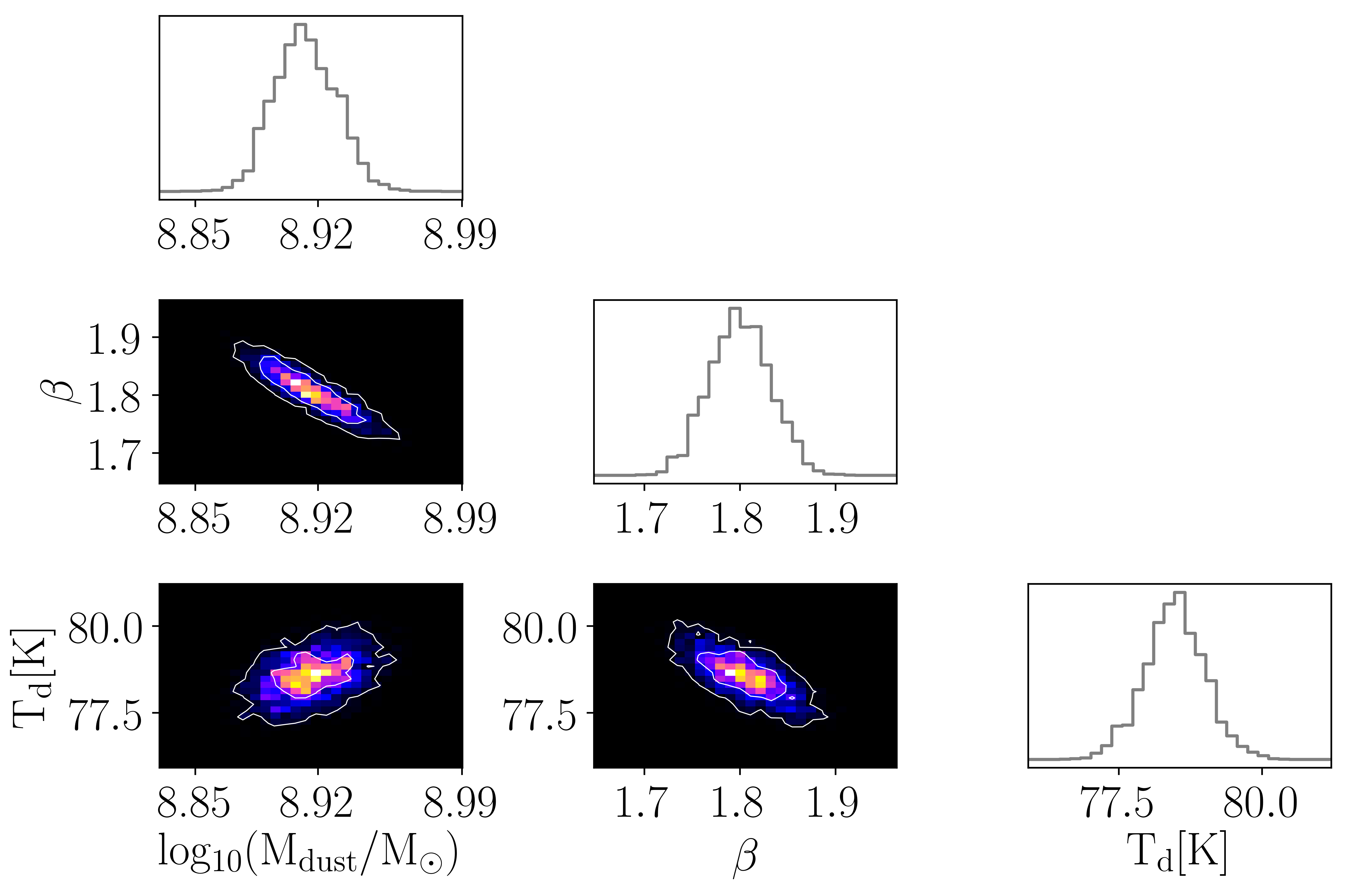}
\caption{The covariance distributions of the three parameters (M$_{dust}$, $\beta$, and T$_{\rm d}$) used to to fit our SED. The top plot of each column is the overall posterior probability distribution for each parameter.}
\label{fig:sedmarg}
\end{figure}

Our best-fit model is shown in Figure \ref{fig:sedfit}, the resulting values are listed in Table \ref{tab:physical}, and the associated covariance plots are shown in Figure \ref{fig:sedmarg}. Fitting a modified blackbody model to our combined dataset yields best-fit values of T$_d=78.6\pm0.5$\,K, $\beta=1.81\pm0.03$, and log($M_{dust}/M_{\odot})=8.92\pm0.02$. This model exhibits a FIR luminosity (i.e., 42.5-122.5\,$\mu$m) of L$_{\rm FIR}=(2.2\pm0.1)\times10^{13}\,$L$_{\odot}$. Using the scaling relation from \citet{kenn98A}, this corresponds to a SFR=$3800\pm100$\,M$_{\odot}$\,year$^{-1}$.

\begin{table} 
\centering
\begin{tabular}{l|c}
T$_{\rm dust}$ [K] & $78.6\pm0.5$\\
$\beta$ & $1.81\pm0.03$\\
log$_{10}($M$_{\rm dust}$/M$_{\odot}$) & $8.92\pm0.02$\\ \hline
L$_{\rm FIR}$ [L$_{\odot}$] & $(2.2\pm0.1)\times10^{13}$\\
SFR [M$_{\odot}$\,year$^{-1}$] & $3800\pm100$\\
\end{tabular}
\caption{Best fit values from a modified blackbody fit to the FIR SED (top) and the resulting FIR luminosity and SFR (below).}
\label{tab:physical}
\end{table}

Using the code CIGALE \citep{noll09}, \citet{ma15} fit a MIR-FIR (without the ALMA, ACT, or ATCA flux densities, but including the high-frequency upper limits) SED of this source, and derived an SFR of $4840^{+1090}_{-890}$\,M$_{\odot}$\,year$^{-1}$. The same study used a derived IR luminosity and the conversions of \citet{kenn98B} and \citet{murp11} to find SFR$_{IR}=3830-5340$\,M$_{\odot}$\,year$^{-1}$. Our value of SFR is only slightly (i.e., $<2\sigma$) lower than these estimates. 

However, our value of $\beta$ is $>3\sigma$ lower than the regularly assumed value of $\beta=2.0$ (\citealt{weis13,gull15,ma15}). Using $\beta=2.0$, previous investigations have yielded dust temperatures of $53\pm5$\,K \citep{weis13} and $52\pm2$\,K \citep{gull15}, which are lower than our value. Since both of these investigations fit an FIR SED with a similar modified blackbody (i.e., with no assumption of optical thinness) to our model \citep{grev12}, this discrepancy in $T_{\rm dust}$ is due to their assumption of $\beta\equiv2.0$.

In order to test whether these previously determined values of $\beta$ and T$_d$ could agree with our results, we fixed each parameter to their literature value, and fit a model using only two or one free variables. In the case where $\beta$ was fixed to the previously-determined value of 2, a nearly identical fit was returned, although with a lower dust mass and temperature, and a worse goodness of fit (i.e., a lower Bayesian evidence). If the dust temperature is instead fixed to 52\,K, then we are unable to fit a reliable model to the data, as all models greatly under-predict the high-frequency flux density. 

We may also also examine the results of erroneously assuming optically thin dust emission. In this case, we find a significantly larger dust mass ($\rm M_{dust}=9.36\pm0.01\,M_{\odot}$), lower dust temperature ($\rm T_{dust}=48.3\pm0.7$\,K), and lower dust emissivity constant ($\beta=1.44\pm0.02$), as well as a comparable FIR luminosity (L$_{\rm FIR}=(2.3\pm0.2)\times10^{13}\,$L$_{\odot}$) and SFR ($4000\pm400$\,M$_{\odot}$\,year$^{-1}$). This low dust temperature is similar to previously determined values, but the best-fit value of $\beta$ is greatly discrepant. This fit underpredicts the high-frequency data ($\lambda_{\rm obs}=160-500\,\mu$m), suggesting a need for an additional MIR power law (e.g., \citealt{case12}). However, we find that our source is not optically thin over the examined frequency range (see Section \ref{DoU}), so this fit is nonphysical.

\subsubsection{Discussion of Uncertainties}\label{DoU}

Here, we discuss possibly detrimental assumptions in our model, as well as the ways in which we attempt to minimize this uncertainty.

\textit{Modified Blackbody: }The primary source of unaccounted uncertainty is the use of of a single-temperature modified blackbody model (MBB). Because of this, our `T$_{\rm dust}$' is a single luminosity-weighted dust temperature, rather than a mass-weighted dust temperature (e.g., \citealt{scov16}) or a distribution of temperatures (e.g., \citealt{zhan18}). This model also assumes no contributions from separate MIR (e.g., \citealt{case12}), radio (e.g., \citealt{yun02}), or higher frequency (e.g., \citealt{leit99}) components. In particular, some studies have found that the addition of a MIR power law to a MBB results in a better fit to the FIR SEDs of some galaxies (e.g., \citealt{ibar15,fais20,reut20})

Even though the MBB function is indeed simple, it has been found to fit dust SEDs very well (e.g., \citealt{bian13,jime18,carn19,croc19,lamp19}). The frequency domain of our model is also controlled to ensure that the contributions from higher and lower frequency components are negligible. If a MIR component was required, then the fit shown in Figure \ref{fig:sedfit} would show a deviation at the high-frequency edge of the dust peak (see figure 1D of \citealt{case12}). Since this frequency range is well-fit, we acknowledge that such a component may be present, but minimal.

\textit{Dust Absorption Coefficient: }While most of the values in our model are allowed to vary, we assume a fixed value for the dust absorption coefficient $\kappa_o$. This practice has been widely adopted for radio-submm SED modelling (e.g., \citealt{wang13, wagg14}), and we adopt a relatively recent dust absorption coefficient (\citealt{beel06}, rather than \citealt{hild83}) that falls into the range of values found by observational studies \citep{alto04}.

\textit{Source Radius}: For the radius of \spt, we average two magnification-corrected effective radii for continuum data taken at $\nu_{\rm rest}\sim666$ and 1000\,GHz ($0.79\pm0.02$\,kpc and $0.73\pm0.03$, respectively; \citealt{apos19}), resulting in a well-constrained value of $r_{eff}=0.76\pm0.02$\,kpc. These continuum observations were taken within the frequency range of our fit, supporting their use.  While our model does not take the uncertainty of $R_{\rm eff}$ into account, we find that perturbing this value by $1\sigma$ results in minimal (i.e., $<3\sigma$) variations in the best-fit values of each parameter.

\textit{Optical Thickness: }As discussed in Section \ref{MD}, we make no assumption on the optical thickness of our source, and therefore use a generalized MBB (similar to e.g., \citealt{leec01,cort20}), rather than one that assumes $\tau_{\nu}<<1$ (e.g., \citealt{carn19,lamp19,vale20}). To test whether this source is optically thin, we apply equation \ref{optth} to the data in Table \ref{tab:sed}, using our best-fit dust mass and emissivity index. We find that $\tau_{\nu}$ ranges from the marginally optically thin regime (0.22 at $\lambda_{obs}=3.0$\,mm) to optically thick (43.99 at $\lambda_{obs}=160$\,$\mu$m), with $\tau_{\nu}\sim1$ at $\lambda_{obs}\sim200$\,$\mu$m. This \textit{a posteriori} calculation verifies that the optically thin assumption cannot be applied to this source.

\textit{Uniformity: }Our model assumes a circular source with a constant radius, luminosity-weighted dust temperature, lensing magnification, and set of dust emission properties (i.e., emission and absorption) across all FIR frequencies. Of course, \spt is not perfectly circular (e.g., \citealt{litk19}) and the magnification factor and intrinsic size of its dust continuum emission vary slightly with frequency (\citealt{apos19}). Specifically, the dust continuum magnification factor has been found to decrease with increasing wavelength: $\mu_{870\mu m}=5.6\pm0.1$, $\mu_{2mm}=5.04\pm0.09$, and $\mu_{3mm}=4.63\pm0.03$ (\citealt{spil16,apos19}). Therefore, our global assumption of $\mu=5.6$ may slightly skew the intrinsic SED. The details of dust heating and emission are also quite complex (e.g., \citealt{drai07}). Therefore, we note that these assumptions have a detrimental effect on the physicality of our best-fit values.

\textit{Fitting Method: }We note that the uncertainties in the best-fit parameter values presented in Table \ref{tab:physical} are simply the standard deviations of the posterior deviations for each parameter as found by MultiNest, and thus do not account for the additional sources of uncertainty discussed here. However, the use of MultiNest allows us to robustly explore the parameter space and to inspect the fits for model degeneracies through the construction of covariance plots.

\subsection{Molecular Hydrogen Mass} 

\subsubsection{HD-Based Estimate}\label{MHM}

Using our estimates of the integrated intensity of HD(J=1-0) emission from this source, we may place limits on the amount of HD and H$_2$ present in \spt. We begin by using our HD(1-0) flux density and equation 2 of \citet{berg13}:
\begin{equation}
F_L=\int S_{\nu}d\nu=\frac{N_{HD}A_{10}h\nu f_u}{4\pi D_L^2}
\end{equation}
where $N_{HD}$ is the number of HD molecules, $A_{10}$ is the Einstein coefficient for spontaneous emission ($5.44\times10^{-8}$\,s$^{-1}$; \citealt{mull05}), \textit{h} is Planck's constant, $\nu$ is the rest frequency of the transition ($2.674986\times10^{12}$\,Hz), $D_L$ is the luminosity distance ($1.70\times10^{27}$\,m), and $f_u$ is the fraction of HD that is in J=1:
\begin{equation}
f_u=\frac{3e^{-128.5/T}}{Q(T)}
\end{equation}
where $T$ is our temperature in Kelvin and Q(T) is the partition function.

Next, we assume that (HD/H$_2$)/(D/H)=1, based on observations of $z\sim2.5$ galaxies with high metallicity and HI column density \citep{lisz15}. We also adopt the primordial value of D/H=$(2.6\pm0.1)\times10^{-5}$ (\citealt{coc04,plan16,novo17}), as the universe is only 1\,Gyr old at $z=5.656$, so the deuterium fraction is still nearly primordial \citep{vand18}. This results in a simple equation for the molecular gas mass:
\begin{equation}
M_{H_2}=\frac{4\pi m_{H_2} D_L^2 F_L}{A_{10}h \nu f_u D/H}
\end{equation}
where $m_{H_2}$ is the mass of a single H$_2$ molecule. Note that while \citet{berg13} considers the total gas mass, we do not consider helium and heavy elements and focus on H$_2$, the main gas component. We assume a gas temperature of $T\sim60$\,K, based on Large Velocity Gradient (LVG) modelling of CO(8-7) and CO(2-1) data \citep{dong19}.

Using our $2\sigma$ upper limit on the integrated flux of HD(1-0) (\rm $\rm F_L<3.8\times10^{-20}\,W\,m^{-2}$), this results in an magnification-corrected upper limit on the molecular gas mass of $\rm M_{H_2}<(6.4\times10^{11})(5.6/\mu)$\,M$_{\odot}$, where $\mu$ is the magnification factor (e.g., \citealt{spil15}). This is in agreement with the CO(2-1) based gas mass of \citet{arav16}: $\rm M_{H_2}=(8.2\pm0.6)\times10^{10}(5.6/\mu)(\alpha_{CO}/0.8)$\,M$_{\odot}$. 

This gas mass estimate may be used to place a constraint on the CO luminosity---H$_2$ mass conversion ratio $\alpha_{CO}=M_{H_2}/L'_{CO}$. We convert the CO(2-1) integrated flux density of \citet{arav16} to a CO(1-0) luminosity using $L'_{CO(2-1)}/L'_{CO(1-0)}=0.85$ (appropriate for starburst galaxies) from table 2 of \citet{cari13}, yielding an observed L'$_{CO(1-0)}=(6.1\pm0.4)\times10^{11}/\mu$\,K\,km\,s$^{-1}$\,pc$^2$, which we then correct for lensing ($\mu=5.6\pm0.1$; \citealt{spil16}). This results in a limit of $\alpha_{CO}<5.8$.

The above values assume that the kinetic temperature of HD is equal to that of CO, as derived by \citet{dong19}. As an alternative estimate of the gas temperature, we briefly consider the possibility that the gas and dust are thermally coupled due to high densities ($n>10^{4.5}$\,cm$^{-3}$; \citealt{gold01}). With this coupling, we may consider the dust temperature derived through our SED fit ($78.6\pm0.05$\,K), resulting in a slightly less conservative upper limit on the gas mass: $\rm M_{H_2}<4.4\times10^{11}\,M_{\odot}$. On the other hand, if the suggested mass-weighted dust temperature of \citet{scov16} is assumed (25\,K), then we find a more conservative gas mass limit of $\rm M_{H_2}<9.9\times10^{12}\,M_{\odot}$.  

\subsubsection{Continuum-Based Estimates}
We may also use our derived dust mass ($10^{9.38}$\,M$_{\odot}$) and a standard gas to dust (GDR) ratio of $\rm \delta_{GDR}=M_{H_2}/M_{dust}=100\times10^{Z-Z_{\odot}}$ \citep{drai07} to determine a separate estimate on M$_{H_2}$. Since no metallicity information is available, we assume sub-solar metallicity for this early galaxy, resulting in  M$_{H_2,GDR}<2.4\times10^{11}$\,M$_{\odot}$.

Alternatively, equation A14 of \citet{scov16} may be applied to our four continuum flux density values that satisfy their $\lambda_{\rm rest}>250$\,$\mu$m criterion ($\lambda_{\rm obs}>1664$\,$\mu$m). Using our derived dust temperature, this yields a magnification-corrected average value of $M_{H_2}=(5.1\pm0.3)\times10^{11}$\,M$_{\odot}$. It should be noted that this  equation was derived using a calibration sample of $z\sim0-3$ galaxies with high stellar masses, and thus high metallicities. Its applicability to higher-redshift objects with possibly lower metallicities like \spt is uncertain. 

These two continuum-based gas mass estimates may also be used to place constraints on the CO luminosity---H$_2$ mass conversion ratio $\alpha_{CO}=M_{H_2}/L'_{CO}$. Both the \citet{scov16} and GDR approaches result in values between the Milky Way (4.6) and starburst (0.8) limits (e.g., \citealt{dadd10}). These gas mass estimates and their resulting $\rm \alpha_{CO}$ values are listed in Table \ref{tab:mh2}. 

\begin{table}
\centering
\begin{tabular}{>{\rowmac}l|>{\rowmac}c|>{\rowmac}c<{\clearrow}}
Method & M$_{\rm H_2}$ [\,M$_{\odot}$] & $\alpha_{\rm CO}$\\ \hline
HD & $<6.4\times10^{11}$ & $<5.8$\\ 
GDR & $<2.4\times10^{11}$ & $<2.2$\\ 
Scoville+16 & $(5.1\pm0.3)\times10^{11}$ & $4.0\pm0.4$\\ \hline 
Aravena+16 & $(8.2\pm0.6)\times10^{10}$  & $\cdots$
\end{tabular}
\caption{Derived molecular gas masses using different estimators, and the corresponding $\alpha_{CO}$ values based on the L'$_{CO(2-1)}$ value of \citet{arav16}. For reference, the CO(2-1)-based gas mass of \citet{arav16} (assuming $\alpha_{CO}=0.8$) is also included.}
\label{tab:mh2}
\end{table}

\section{Conclusions}
In this work, we have presented the first upper limit on the luminosity of HD(J=1-0) in \spt, fitted a modified blackbody model to a FIR SED of the source, and derived several estimates of the mass of molecular gas in the system. All of these results confirm the extreme starburst nature of this galaxy, but there is variation in some derived values. 

No significant HD(J=1-0) emission is detected, implying a molecular gas mass of $\rm M_{H_2}<6.4\times10^{11}\,M_{\odot}$. This estimate is in agreement with a previous estimate of $\rm M_{H_2}=(8.2\pm0.6)\times10^{10}\,M_{\odot}$ based on CO(2-1) emission \citep{arav16}. 

Using archival ALMA observations, we are able to fill in the red side of the dust SED, resulting in a better constraint in the dust emissivity spectral index $\beta$ and new estimates on the dust mass and dust temperature. Our best-fit luminosity-weighted dust temperature ($T=78.6\pm0.5$\,K) is greater than both the extreme mass-weighted temperature case of $25$\,K \citep{scov16} and the previous estimate of $53$\,K \citep{weis13}. This discrepancy is possibly due to the extreme optical depth of our source (i.e., $\tau_{o}$ reaches 43.99 at $\lambda_{obs}=160\,\mu$m), or our different method of deriving the dust emissivity spectral index $\beta$. Our derived value of $\beta=1.81\pm0.03$ is reasonable considering the standard range of $\beta=1.5-2.0$ for high redshift, dusty galaxies (e.g., \citealt{chap05,chap11,case11}. We present a thorough discussion of the assumptions of our models, and state possible sources of unstated uncertainty in our best-fit models.

The molecular mass of the entire system was also estimated with two continuum-based methods. First, the \citet{scov16} S($\nu_{rest}<1.2$\,THz) estimator was applied to four of our SED values, resulting in an average value of $\rm M_{H_2}=(5.1\pm0.3)\times10^{11}$\,M$_{\odot}$. In addition, our dust mass was converted to a gas mass by assuming sub-solar metallicity and a possible dust-to-gas ratio, yielding an upper limit of $\rm M_{H_2}<2.4\times10^{11}$\,M$_{\odot}$. The difference between these values may suggest that the dust temperature is underestimated, the galaxy has a high metallicity, or simply that this source does not follow the \citet{scov16} relation, which was created using local starbursts.

These three gas mass estimates may be used to place constraints on $\alpha_{\rm CO}$, resulting in $<5.8$, $<2.2$, and $4.0\pm0.4$ for the HD, GDR, and \citet{scov16} approaches, respectively. While they are disparate, they either agree with or are between the MW-like $\alpha_{\rm CO}\sim4.6$ (e.g., \citealt{dadd10}) and the starburst $\alpha_{\rm CO}\sim0.8$, which was assumed by \citet{arav16}. Due to the large star formation rate of \spt ($\sim4000$\,M$_{\odot}$\,year$^{-1}$), these high conversion factors may be surprising. However, a recent investigation of the submillimeter galaxy AzTEC/C159 by \citet{jime18} found $\alpha_{CO}\sim4$, despite the high star formation rate of the source ($\sim700$\,M$_{\odot}$\,year$^{-1}$). One of the interpretations of this results was that AzTEC/C159 is in the early stages of a starburst. It is thus possible that \spt is undergoing a massive starburst driven by the presence of a large amount of pristine gas.

Due to the lensed nature of the source, we are unable to present any exact SFR or molecular mass surface densities. However, as a test of our values, we consider a size of $\sim0.76$\,kpc \citep{apos19} and a maximal limit of $\rm A_{SFR}=A_{H_2}$. This would suggest a $\Sigma_{SFR}\sim2\times10^3$\,M$_{\odot}$\,year$^{-1}$\,kpc$^{-2}$ and $\Sigma_{H2}\sim5.5\times10^4$\,M$_{\odot}$\,pc$^{-2}$. This places \spt squarely in the starburst region of a Kennicutt-Schmidt diagram (e.g., \citealt{kenn98A}). These results confirm that \spt is a highly starbursting, dusty, extremely luminous galaxy.

We note that our observations were designed to maximize sensitivity to compact emission, and so the nondetection of HD(1-0) may be partly due to the extended nature of the emission. In addition, the expected  line is coincident with a series of atmospheric transmission lines, resulting in a nonconstant noise level. Future observations of HD(1-0) may benefit from targeting sources at higher-redshift (i.e., $z>7.5$), where the line will fall into a frequency range of higher atmospheric transmission (i.e., ALMA band 7 or below). With the rising number detections of CO emission at $z>5$ (e.g., \citealt{dodo18,pave18,pave19,riec20}), it will be of interest to have an independent tracer of molecular gas mass in the early Universe.

\section*{Data Availability}
The data analyzed in this work are available from the ALMA data archive (https://almascience.nrao.edu/asax/) under project code 2016.1.01313.S (Band 8).

\section*{Acknowledgements}
This paper makes use of the following ALMA data: 2015.1.01580.S, 2013.1.01231.S, 2016.1.00654.S, and 2016.1.01313.S.
ALMA is a partnership of ESO (representing its member states), NSF (USA) and NINS (Japan), together with NRC (Canada), MOST and ASIAA (Taiwan), and KASI (Republic of Korea), in cooperation with the Republic of Chile. The Joint ALMA Observatory is operated by ESO, AUI/NRAO and NAOJ. G.C.J. and R.M. acknowledge ERC Advanced Grant 695671 ``QUENCH'' and support by the Science and Technology Facilities Council (STFC). S.C. acknowledges support from the ERC Advanced Grant INTERSTELLAR H2020/740120. We thank the anonymous referee for constructive feedback that strengthened this manuscript.

\bibliographystyle{mnras}
\bibliography{spthd}

\begin{thebibliography}{}
\makeatletter
\relax
\def\mn@urlcharsother{\let\do\@makeother \do\$\do\&\do\#\do\^\do\_\do\%\do\~}
\def\mn@doi{\begingroup\mn@urlcharsother \@ifnextchar [ {\mn@doi@}
  {\mn@doi@[]}}
\def\mn@doi@[#1]#2{\def\@tempa{#1}\ifx\@tempa\@empty \href
  {http://dx.doi.org/#2} {doi:#2}\else \href {http://dx.doi.org/#2} {#1}\fi
  \endgroup}
\def\mn@eprint#1#2{\mn@eprint@#1:#2::\@nil}
\def\mn@eprint@arXiv#1{\href {http://arxiv.org/abs/#1} {{\tt arXiv:#1}}}
\def\mn@eprint@dblp#1{\href {http://dblp.uni-trier.de/rec/bibtex/#1.xml}
  {dblp:#1}}
\def\mn@eprint@#1:#2:#3:#4\@nil{\def\@tempa {#1}\def\@tempb {#2}\def\@tempc
  {#3}\ifx \@tempc \@empty \let \@tempc \@tempb \let \@tempb \@tempa \fi \ifx
  \@tempb \@empty \def\@tempb {arXiv}\fi \@ifundefined
  {mn@eprint@\@tempb}{\@tempb:\@tempc}{\expandafter \expandafter \csname
  mn@eprint@\@tempb\endcsname \expandafter{\@tempc}}}

\bibitem[\protect\citeauthoryear{{Alton}, {Xilouris}, {Misiriotis}, {Dasyra}
  \& {Dumke}}{{Alton} et~al.}{2004}]{alto04}
{Alton} P.~B.,  {Xilouris} E.~M.,  {Misiriotis} A.,  {Dasyra} K.~M.,   {Dumke}
  M.,  2004, \mn@doi [\aap] {10.1051/0004-6361:20040438}, \href
  {https://ui.adsabs.harvard.edu/abs/2004A&A...425..109A} {425, 109}

\bibitem[\protect\citeauthoryear{{Apostolovski} et~al.,}{{Apostolovski}
  et~al.}{2019}]{apos19}
{Apostolovski} Y.,  et~al., 2019, \mn@doi [\aap] {10.1051/0004-6361/201935308},
  \href {https://ui.adsabs.harvard.edu/abs/2019A&A...628A..23A} {628, A23}

\bibitem[\protect\citeauthoryear{{Aravena} et~al.,}{{Aravena}
  et~al.}{2016}]{arav16}
{Aravena} M.,  et~al., 2016, \mn@doi [\mnras] {10.1093/mnras/stw275}, \href
  {http://adsabs.harvard.edu/abs/2016MNRAS.457.4406A} {457, 4406}

\bibitem[\protect\citeauthoryear{{Balashev}, {Ivanchik}  \&
  {Varshalovich}}{{Balashev} et~al.}{2010}]{bala10}
{Balashev} S.~A.,  {Ivanchik} A.~V.,   {Varshalovich} D.~A.,  2010, \mn@doi
  [Astronomy Letters] {10.1134/S1063773710110010}, \href
  {http://adsabs.harvard.edu/abs/2010AstL...36..761B} {36, 761}

\bibitem[\protect\citeauthoryear{{Beelen}, {Cox}, {Benford}, {Dowell},
  {Kov{\'a}cs}, {Bertoldi}, {Omont}  \& {Carilli}}{{Beelen}
  et~al.}{2006}]{beel06}
{Beelen} A.,  {Cox} P.,  {Benford} D.~J.,  {Dowell} C.~D.,  {Kov{\'a}cs} A.,
  {Bertoldi} F.,  {Omont} A.,   {Carilli} C.~L.,  2006, \mn@doi [\apj]
  {10.1086/500636}, \href
  {https://ui.adsabs.harvard.edu/abs/2006ApJ...642..694B} {642, 694}

\bibitem[\protect\citeauthoryear{{Bergin} et~al.,}{{Bergin}
  et~al.}{2013}]{berg13}
{Bergin} E.~A.,  et~al., 2013, \mn@doi [\nat] {10.1038/nature11805}, \href
  {http://adsabs.harvard.edu/abs/2013Natur.493..644B} {493, 644}

\bibitem[\protect\citeauthoryear{{Bianchi}}{{Bianchi}}{2013}]{bian13}
{Bianchi} S.,  2013, \mn@doi [\aap] {10.1051/0004-6361/201220866}, \href
  {https://ui.adsabs.harvard.edu/abs/2013A&A...552A..89B} {552, A89}

\bibitem[\protect\citeauthoryear{{Bolatto}, {Wolfire}  \& {Leroy}}{{Bolatto}
  et~al.}{2013}]{bola13}
{Bolatto} A.~D.,  {Wolfire} M.,   {Leroy} A.~K.,  2013, \mn@doi [\araa]
  {10.1146/annurev-astro-082812-140944}, \href
  {http://adsabs.harvard.edu/abs/2013ARA%26A..51..207B} {51, 207}

\bibitem[\protect\citeauthoryear{{Buchner} et~al.,}{{Buchner}
  et~al.}{2014}]{buch14}
{Buchner} J.,  et~al., 2014, \mn@doi [\aap] {10.1051/0004-6361/201322971},
  \href {http://adsabs.harvard.edu/abs/2014A%26A...564A.125B} {564, A125}

\bibitem[\protect\citeauthoryear{{Carilli} \& {Walter}}{{Carilli} \&
  {Walter}}{2013}]{cari13}
{Carilli} C.~L.,  {Walter} F.,  2013, \mn@doi [\araa]
  {10.1146/annurev-astro-082812-140953}, \href
  {https://ui.adsabs.harvard.edu/abs/2013ARA&A..51..105C} {51, 105}

\bibitem[\protect\citeauthoryear{{Carniani} et~al.,}{{Carniani}
  et~al.}{2019}]{carn19}
{Carniani} S.,  et~al., 2019, \mn@doi [\mnras] {10.1093/mnras/stz2410}, \href
  {https://ui.adsabs.harvard.edu/abs/2019MNRAS.489.3939C} {489, 3939}

\bibitem[\protect\citeauthoryear{{Casey}}{{Casey}}{2012}]{case12}
{Casey} C.~M.,  2012, \mn@doi [\mnras] {10.1111/j.1365-2966.2012.21455.x},
  \href {http://adsabs.harvard.edu/abs/2012MNRAS.425.3094C} {425, 3094}

\bibitem[\protect\citeauthoryear{{Casey} et~al.,}{{Casey}
  et~al.}{2011}]{case11}
{Casey} C.~M.,  et~al., 2011, \mn@doi [\mnras]
  {10.1111/j.1365-2966.2011.18885.x}, \href
  {http://adsabs.harvard.edu/abs/2011MNRAS.415.2723C} {415, 2723}

\bibitem[\protect\citeauthoryear{{Cazaux} \& {Spaans}}{{Cazaux} \&
  {Spaans}}{2009}]{caza09}
{Cazaux} S.,  {Spaans} M.,  2009, \mn@doi [\aap] {10.1051/0004-6361:200811302},
  \href {http://adsabs.harvard.edu/abs/2009A%26A...496..365C} {496, 365}

\bibitem[\protect\citeauthoryear{{Chapin} et~al.,}{{Chapin}
  et~al.}{2011}]{chap11}
{Chapin} E.~L.,  et~al., 2011, \mn@doi [\mnras]
  {10.1111/j.1365-2966.2010.17697.x}, \href
  {http://adsabs.harvard.edu/abs/2011MNRAS.411..505C} {411, 505}

\bibitem[\protect\citeauthoryear{{Chapman}, {Blain}, {Smail}  \&
  {Ivison}}{{Chapman} et~al.}{2005}]{chap05}
{Chapman} S.~C.,  {Blain} A.~W.,  {Smail} I.,   {Ivison} R.~J.,  2005, \mn@doi
  [\apj] {10.1086/428082}, \href
  {http://adsabs.harvard.edu/abs/2005ApJ...622..772C} {622, 772}

\bibitem[\protect\citeauthoryear{{Coc}, {Vangioni-Flam}, {Descouvemont},
  {Adahchour}  \& {Angulo}}{{Coc} et~al.}{2004}]{coc04}
{Coc} A.,  {Vangioni-Flam} E.,  {Descouvemont} P.,  {Adahchour} A.,   {Angulo}
  C.,  2004, \mn@doi [\apj] {10.1086/380121}, \href
  {http://adsabs.harvard.edu/abs/2004ApJ...600..544C} {600, 544}

\bibitem[\protect\citeauthoryear{{Cortzen} et~al.,}{{Cortzen}
  et~al.}{2020}]{cort20}
{Cortzen} I.,  et~al., 2020, \mn@doi [\aap] {10.1051/0004-6361/201937217},
  \href {https://ui.adsabs.harvard.edu/abs/2020A&A...634L..14C} {634, L14}

\bibitem[\protect\citeauthoryear{{Crocker} et~al.,}{{Crocker}
  et~al.}{2019}]{croc19}
{Crocker} A.~F.,  et~al., 2019, \mn@doi [\apj] {10.3847/1538-4357/ab4196},
  \href {https://ui.adsabs.harvard.edu/abs/2019ApJ...887..105C} {887, 105}

\bibitem[\protect\citeauthoryear{{D'Odorico} et~al.,}{{D'Odorico}
  et~al.}{2018}]{dodo18}
{D'Odorico} V.,  et~al., 2018, \mn@doi [\apjl] {10.3847/2041-8213/aad7b7},
  \href {https://ui.adsabs.harvard.edu/abs/2018ApJ...863L..29D} {863, L29}

\bibitem[\protect\citeauthoryear{{Daddi} et~al.,}{{Daddi}
  et~al.}{2010}]{dadd10}
{Daddi} E.,  et~al., 2010, \mn@doi [\apjl] {10.1088/2041-8205/714/1/L118},
  \href {https://ui.adsabs.harvard.edu/abs/2010ApJ...714L.118D} {714, L118}

\bibitem[\protect\citeauthoryear{{Dapr{\`a}}, {van der Laan}, {Murphy}  \&
  {Ubachs}}{{Dapr{\`a}} et~al.}{2017}]{dapr17}
{Dapr{\`a}} M.,  {van der Laan} M.,  {Murphy} M.~T.,   {Ubachs} W.,  2017,
  \mn@doi [\mnras] {10.1093/mnras/stw3003}, \href
  {http://adsabs.harvard.edu/abs/2017MNRAS.465.4057D} {465, 4057}

\bibitem[\protect\citeauthoryear{{Dong} et~al.,}{{Dong} et~al.}{2019}]{dong19}
{Dong} C.,  et~al., 2019, \mn@doi [\apj] {10.3847/1538-4357/ab02fe}, \href
  {https://ui.adsabs.harvard.edu/abs/2019ApJ...873...50D} {873, 50}

\bibitem[\protect\citeauthoryear{{Draine} et~al.,}{{Draine}
  et~al.}{2007}]{drai07}
{Draine} B.~T.,  et~al., 2007, \mn@doi [\apj] {10.1086/518306}, \href
  {http://adsabs.harvard.edu/abs/2007ApJ...663..866D} {663, 866}

\bibitem[\protect\citeauthoryear{{Faisst}, {Fudamoto}, {Oesch}, {Scoville},
  {Riechers}, {Pavesi}  \& {Capak}}{{Faisst} et~al.}{2020}]{fais20}
{Faisst} A.~L.,  {Fudamoto} Y.,  {Oesch} P.~A.,  {Scoville} N.,  {Riechers}
  D.~A.,  {Pavesi} R.,   {Capak} P.,  2020, arXiv e-prints, \href
  {https://ui.adsabs.harvard.edu/abs/2020arXiv200507716F} {p. arXiv:2005.07716}

\bibitem[\protect\citeauthoryear{{Feroz} \& {Hobson}}{{Feroz} \&
  {Hobson}}{2008}]{fero08}
{Feroz} F.,  {Hobson} M.~P.,  2008, \mn@doi [\mnras]
  {10.1111/j.1365-2966.2007.12353.x}, \href
  {http://adsabs.harvard.edu/abs/2008MNRAS.384..449F} {384, 449}

\bibitem[\protect\citeauthoryear{{Goldsmith}}{{Goldsmith}}{2001}]{gold01}
{Goldsmith} P.~F.,  2001, \mn@doi [\apj] {10.1086/322255}, \href
  {https://ui.adsabs.harvard.edu/abs/2001ApJ...557..736G} {557, 736}

\bibitem[\protect\citeauthoryear{{Greve} et~al.,}{{Greve}
  et~al.}{2012}]{grev12}
{Greve} T.~R.,  et~al., 2012, \mn@doi [\apj] {10.1088/0004-637X/756/1/101},
  \href {https://ui.adsabs.harvard.edu/abs/2012ApJ...756..101G} {756, 101}

\bibitem[\protect\citeauthoryear{{Gullberg} et~al.,}{{Gullberg}
  et~al.}{2015}]{gull15}
{Gullberg} B.,  et~al., 2015, \mn@doi [\mnras] {10.1093/mnras/stv372}, \href
  {http://adsabs.harvard.edu/abs/2015MNRAS.449.2883G} {449, 2883}

\bibitem[\protect\citeauthoryear{{Hildebrand}}{{Hildebrand}}{1983}]{hild83}
{Hildebrand} R.~H.,  1983, \qjras, \href
  {https://ui.adsabs.harvard.edu/abs/1983QJRAS..24..267H} {24, 267}

\bibitem[\protect\citeauthoryear{{Ibar} et~al.,}{{Ibar} et~al.}{2015}]{ibar15}
{Ibar} E.,  et~al., 2015, \mn@doi [\mnras] {10.1093/mnras/stv439}, \href
  {https://ui.adsabs.harvard.edu/abs/2015MNRAS.449.2498I} {449, 2498}

\bibitem[\protect\citeauthoryear{{Ivanchik}, {Petitjean}, {Balashev},
  {Srianand}, {Varshalovich}, {Ledoux}  \& {Noterdaeme}}{{Ivanchik}
  et~al.}{2010}]{ivan10}
{Ivanchik} A.~V.,  {Petitjean} P.,  {Balashev} S.~A.,  {Srianand} R.,
  {Varshalovich} D.~A.,  {Ledoux} C.,   {Noterdaeme} P.,  2010, \mn@doi
  [\mnras] {10.1111/j.1365-2966.2010.16382.x}, \href
  {http://adsabs.harvard.edu/abs/2010MNRAS.404.1583I} {404, 1583}

\bibitem[\protect\citeauthoryear{{Jansen}, {van Dishoeck}, {Black}, {Spaans}
  \& {Sosin}}{{Jansen} et~al.}{1995}]{jans95}
{Jansen} D.~J.,  {van Dishoeck} E.~F.,  {Black} J.~H.,  {Spaans} M.,   {Sosin}
  C.,  1995, \aap, \href
  {https://ui.adsabs.harvard.edu/abs/1995A&A...302..223J} {302, 223}

\bibitem[\protect\citeauthoryear{{Jim{\'e}nez-Andrade}
  et~al.,}{{Jim{\'e}nez-Andrade} et~al.}{2018}]{jime18}
{Jim{\'e}nez-Andrade} E.~F.,  et~al., 2018, \mn@doi [\aap]
  {10.1051/0004-6361/201732186}, \href
  {http://adsabs.harvard.edu/abs/2018A%26A...615A..25J} {615, A25}

\bibitem[\protect\citeauthoryear{{Jones}, {Maiolino}, {Caselli}  \&
  {Carniani}}{{Jones} et~al.}{2019}]{jone19}
{Jones} G.~C.,  {Maiolino} R.,  {Caselli} P.,   {Carniani} S.,  2019, \mn@doi
  [\aap] {10.1051/0004-6361/201936989}, \href
  {https://ui.adsabs.harvard.edu/abs/2019A%26A...632L...7J} {632, J7}

\bibitem[\protect\citeauthoryear{{Kennicutt}}{{Kennicutt}}{1998a}]{kenn98B}
{Kennicutt} Jr. R.~C.,  1998a, \mn@doi [\araa]
  {10.1146/annurev.astro.36.1.189}, \href
  {http://adsabs.harvard.edu/abs/1998ARA%26A..36..189K} {36, 189}

\bibitem[\protect\citeauthoryear{{Kennicutt}}{{Kennicutt}}{1998b}]{kenn98A}
{Kennicutt} Jr. R.~C.,  1998b, \mn@doi [\apj] {10.1086/305588}, \href
  {http://adsabs.harvard.edu/abs/1998ApJ...498..541K} {498, 541}

\bibitem[\protect\citeauthoryear{{Kennicutt} \& {Evans}}{{Kennicutt} \&
  {Evans}}{2012}]{kenn12}
{Kennicutt} R.~C.,  {Evans} N.~J.,  2012, \mn@doi [\araa]
  {10.1146/annurev-astro-081811-125610}, \href
  {http://adsabs.harvard.edu/abs/2012ARA%26A..50..531K} {50, 531}

\bibitem[\protect\citeauthoryear{{Lacour} et~al.,}{{Lacour}
  et~al.}{2005}]{laco05}
{Lacour} S.,  et~al., 2005, \mn@doi [\aap] {10.1051/0004-6361:20041589}, \href
  {http://adsabs.harvard.edu/abs/2005A%26A...430..967L} {430, 967}

\bibitem[\protect\citeauthoryear{{Lamperti} et~al.,}{{Lamperti}
  et~al.}{2019}]{lamp19}
{Lamperti} I.,  et~al., 2019, \mn@doi [\mnras] {10.1093/mnras/stz2311}, \href
  {https://ui.adsabs.harvard.edu/abs/2019MNRAS.489.4389L} {489, 4389}

\bibitem[\protect\citeauthoryear{{Leech}, {Metcalfe}  \& {Altieri}}{{Leech}
  et~al.}{2001}]{leec01}
{Leech} K.~J.,  {Metcalfe} L.,   {Altieri} B.,  2001, \mn@doi [\mnras]
  {10.1046/j.1365-8711.2001.04941.x}, \href
  {https://ui.adsabs.harvard.edu/abs/2001MNRAS.328.1125L} {328, 1125}

\bibitem[\protect\citeauthoryear{{Leitherer} et~al.,}{{Leitherer}
  et~al.}{1999}]{leit99}
{Leitherer} C.,  et~al., 1999, \mn@doi [\apjs] {10.1086/313233}, \href
  {https://ui.adsabs.harvard.edu/abs/1999ApJS..123....3L} {123, 3}

\bibitem[\protect\citeauthoryear{{Liszt}}{{Liszt}}{2015}]{lisz15}
{Liszt} H.~S.,  2015, \mn@doi [\apj] {10.1088/0004-637X/799/1/66}, \href
  {http://adsabs.harvard.edu/abs/2015ApJ...799...66L} {799, 66}

\bibitem[\protect\citeauthoryear{{Litke} et~al.,}{{Litke}
  et~al.}{2019}]{litk19}
{Litke} K.~C.,  et~al., 2019, \mn@doi [\apj] {10.3847/1538-4357/aaf057}, \href
  {http://adsabs.harvard.edu/abs/2019ApJ...870...80L} {870, 80}

\bibitem[\protect\citeauthoryear{{Ma} et~al.,}{{Ma} et~al.}{2015}]{ma15}
{Ma} J.,  et~al., 2015, \mn@doi [\apj] {10.1088/0004-637X/812/1/88}, \href
  {http://adsabs.harvard.edu/abs/2015ApJ...812...88M} {812, 88}

\bibitem[\protect\citeauthoryear{{Ma} et~al.,}{{Ma} et~al.}{2016}]{ma16}
{Ma} J.,  et~al., 2016, \mn@doi [\apj] {10.3847/0004-637X/832/2/114}, \href
  {http://adsabs.harvard.edu/abs/2016ApJ...832..114M} {832, 114}

\bibitem[\protect\citeauthoryear{{Marsden} et~al.,}{{Marsden}
  et~al.}{2014}]{mars14}
{Marsden} D.,  et~al., 2014, \mn@doi [\mnras] {10.1093/mnras/stu001}, \href
  {http://adsabs.harvard.edu/abs/2014MNRAS.439.1556M} {439, 1556}

\bibitem[\protect\citeauthoryear{{McClure} et~al.,}{{McClure}
  et~al.}{2016}]{mccl16}
{McClure} M.~K.,  et~al., 2016, \mn@doi [\apj] {10.3847/0004-637X/831/2/167},
  \href {https://ui.adsabs.harvard.edu/abs/2016ApJ...831..167M} {831, 167}

\bibitem[\protect\citeauthoryear{{M{\"u}ller}, {Schl{\"o}der}, {Stutzki}  \&
  {Winnewisser}}{{M{\"u}ller} et~al.}{2005}]{mull05}
{M{\"u}ller} H.~S.~P.,  {Schl{\"o}der} F.,  {Stutzki} J.,   {Winnewisser} G.,
  2005, \mn@doi [Journal of Molecular Structure]
  {10.1016/j.molstruc.2005.01.027}, \href
  {http://adsabs.harvard.edu/abs/2005JMoSt.742..215M} {742, 215}

\bibitem[\protect\citeauthoryear{{Murphy} et~al.,}{{Murphy}
  et~al.}{2011}]{murp11}
{Murphy} E.~J.,  et~al., 2011, \mn@doi [\apj] {10.1088/0004-637X/737/2/67},
  \href {http://adsabs.harvard.edu/abs/2011ApJ...737...67M} {737, 67}

\bibitem[\protect\citeauthoryear{{Narayanan}, {Krumholz}, {Ostriker}  \&
  {Hernquist}}{{Narayanan} et~al.}{2012}]{nara12}
{Narayanan} D.,  {Krumholz} M.~R.,  {Ostriker} E.~C.,   {Hernquist} L.,  2012,
  \mn@doi [\mnras] {10.1111/j.1365-2966.2012.20536.x}, \href
  {http://adsabs.harvard.edu/abs/2012MNRAS.421.3127N} {421, 3127}

\bibitem[\protect\citeauthoryear{{Neufeld} et~al.,}{{Neufeld}
  et~al.}{2006}]{neuf06}
{Neufeld} D.~A.,  et~al., 2006, \mn@doi [\apjl] {10.1086/507130}, \href
  {http://adsabs.harvard.edu/abs/2006ApJ...647L..33N} {647, L33}

\bibitem[\protect\citeauthoryear{{Noll}, {Burgarella}, {Giovannoli}, {Buat},
  {Marcillac}  \& {Mu{\~n}oz-Mateos}}{{Noll} et~al.}{2009}]{noll09}
{Noll} S.,  {Burgarella} D.,  {Giovannoli} E.,  {Buat} V.,  {Marcillac} D.,
  {Mu{\~n}oz-Mateos} J.~C.,  2009, \mn@doi [\aap]
  {10.1051/0004-6361/200912497}, \href
  {http://adsabs.harvard.edu/abs/2009A%26A...507.1793N} {507, 1793}

\bibitem[\protect\citeauthoryear{{Noterdaeme}, {Petitjean}, {Ledoux},
  {Srianand}  \& {Ivanchik}}{{Noterdaeme} et~al.}{2008}]{note08}
{Noterdaeme} P.,  {Petitjean} P.,  {Ledoux} C.,  {Srianand} R.,   {Ivanchik}
  A.,  2008, \mn@doi [\aap] {10.1051/0004-6361:200810414}, \href
  {http://adsabs.harvard.edu/abs/2008A%26A...491..397N} {491, 397}

\bibitem[\protect\citeauthoryear{{Noterdaeme}, {Petitjean}, {Ledoux},
  {L{\'o}pez}, {Srianand}  \& {Vergani}}{{Noterdaeme} et~al.}{2010}]{note10}
{Noterdaeme} P.,  {Petitjean} P.,  {Ledoux} C.,  {L{\'o}pez} S.,  {Srianand}
  R.,   {Vergani} S.~D.,  2010, \mn@doi [\aap] {10.1051/0004-6361/201015147},
  \href {http://adsabs.harvard.edu/abs/2010A%26A...523A..80N} {523, A80}

\bibitem[\protect\citeauthoryear{{Novosyadlyj}, {Sergijenko}  \&
  {Shulga}}{{Novosyadlyj} et~al.}{2017}]{novo17}
{Novosyadlyj} B.,  {Sergijenko} O.,   {Shulga} V.~M.,  2017, \mn@doi
  [Kinematics and Physics of Celestial Bodies] {10.3103/S088459131706006X},
  \href {http://adsabs.harvard.edu/abs/2017KPCB...33..255N} {33, 255}

\bibitem[\protect\citeauthoryear{{Oliveira}, {Sembach}, {Tumlinson}, {O'Meara}
  \& {Thom}}{{Oliveira} et~al.}{2014}]{oliv14}
{Oliveira} C.~M.,  {Sembach} K.~R.,  {Tumlinson} J.,  {O'Meara} J.,   {Thom}
  C.,  2014, \mn@doi [\apj] {10.1088/0004-637X/783/1/22}, \href
  {http://adsabs.harvard.edu/abs/2014ApJ...783...22O} {783, 22}

\bibitem[\protect\citeauthoryear{{Pavesi} et~al.,}{{Pavesi}
  et~al.}{2018}]{pave18}
{Pavesi} R.,  et~al., 2018, \mn@doi [\apj] {10.3847/1538-4357/aac6b6}, \href
  {https://ui.adsabs.harvard.edu/abs/2018ApJ...861...43P} {861, 43}

\bibitem[\protect\citeauthoryear{{Pavesi}, {Riechers}, {Faisst}, {Stacey}  \&
  {Capak}}{{Pavesi} et~al.}{2019}]{pave19}
{Pavesi} R.,  {Riechers} D.~A.,  {Faisst} A.~L.,  {Stacey} G.~J.,   {Capak}
  P.~L.,  2019, \mn@doi [\apj] {10.3847/1538-4357/ab3a46}, \href
  {https://ui.adsabs.harvard.edu/abs/2019ApJ...882..168P} {882, 168}

\bibitem[\protect\citeauthoryear{{Planck Collaboration} et~al.,}{{Planck
  Collaboration} et~al.}{2016}]{plan16}
{Planck Collaboration} et~al., 2016, \mn@doi [\aap]
  {10.1051/0004-6361/201525830}, \href
  {http://adsabs.harvard.edu/abs/2016A%26A...594A..13P} {594, A13}

\bibitem[\protect\citeauthoryear{{Reuter} et~al.,}{{Reuter}
  et~al.}{2020}]{reut20}
{Reuter} C.,  et~al., 2020, arXiv e-prints, \href
  {https://ui.adsabs.harvard.edu/abs/2020arXiv200614060R} {p. arXiv:2006.14060}

\bibitem[\protect\citeauthoryear{{Riechers} et~al.,}{{Riechers}
  et~al.}{2020}]{riec20}
{Riechers} D.~A.,  et~al., 2020, \mn@doi [\apj] {10.3847/1538-4357/ab8c48},
  \href {https://ui.adsabs.harvard.edu/abs/2020ApJ...895...81R} {895, 81}

\bibitem[\protect\citeauthoryear{{Saintonge} et~al.,}{{Saintonge}
  et~al.}{2013}]{sain13}
{Saintonge} A.,  et~al., 2013, \mn@doi [\apj] {10.1088/0004-637X/778/1/2},
  \href {http://adsabs.harvard.edu/abs/2013ApJ...778....2S} {778, 2}

\bibitem[\protect\citeauthoryear{{Scoville} et~al.,}{{Scoville}
  et~al.}{2016}]{scov16}
{Scoville} N.,  et~al., 2016, \mn@doi [\apj] {10.3847/0004-637X/820/2/83},
  \href {http://adsabs.harvard.edu/abs/2016ApJ...820...83S} {820, 83}

\bibitem[\protect\citeauthoryear{{Spilker} et~al.,}{{Spilker}
  et~al.}{2015}]{spil15}
{Spilker} J.~S.,  et~al., 2015, \mn@doi [\apj] {10.1088/0004-637X/811/2/124},
  \href {http://adsabs.harvard.edu/abs/2015ApJ...811..124S} {811, 124}

\bibitem[\protect\citeauthoryear{{Spilker} et~al.,}{{Spilker}
  et~al.}{2016}]{spil16}
{Spilker} J.~S.,  et~al., 2016, \mn@doi [\apj] {10.3847/0004-637X/826/2/112},
  \href {http://adsabs.harvard.edu/abs/2016ApJ...826..112S} {826, 112}

\bibitem[\protect\citeauthoryear{{Spitzer}, {Drake}, {Jenkins}, {Morton},
  {Rogerson}  \& {York}}{{Spitzer} et~al.}{1973}]{spit73}
{Spitzer} L.,  {Drake} J.~F.,  {Jenkins} E.~B.,  {Morton} D.~C.,  {Rogerson}
  J.~B.,   {York} D.~G.,  1973, \mn@doi [\apjl] {10.1086/181197}, \href
  {http://adsabs.harvard.edu/abs/1973ApJ...181L.116S} {181, L116}

\bibitem[\protect\citeauthoryear{{Tielens}}{{Tielens}}{2005}]{tiel05}
{Tielens} A.~G.~G.~M.,  2005, {The Physics and Chemistry of the Interstellar
  Medium}

\bibitem[\protect\citeauthoryear{{Valentino} et~al.,}{{Valentino}
  et~al.}{2020}]{vale20}
{Valentino} F.,  et~al., 2020, \mn@doi [\apj] {10.3847/1538-4357/ab6603}, \href
  {https://ui.adsabs.harvard.edu/abs/2020ApJ...890...24V} {890, 24}

\bibitem[\protect\citeauthoryear{{Vieira} et~al.,}{{Vieira}
  et~al.}{2013}]{viei13}
{Vieira} J.~D.,  et~al., 2013, \mn@doi [\nat] {10.1038/nature12001}, \href
  {http://adsabs.harvard.edu/abs/2013Natur.495..344V} {495, 344}

\bibitem[\protect\citeauthoryear{{Wagg} et~al.,}{{Wagg} et~al.}{2014}]{wagg14}
{Wagg} J.,  et~al., 2014, \mn@doi [\apj] {10.1088/0004-637X/783/2/71}, \href
  {http://adsabs.harvard.edu/abs/2014ApJ...783...71W} {783, 71}

\bibitem[\protect\citeauthoryear{{Wang} et~al.,}{{Wang} et~al.}{2013}]{wang13}
{Wang} R.,  et~al., 2013, \mn@doi [\apj] {10.1088/0004-637X/773/1/44}, \href
  {https://ui.adsabs.harvard.edu/abs/2013ApJ...773...44W} {773, 44}

\bibitem[\protect\citeauthoryear{{Watson}}{{Watson}}{1973}]{wats73}
{Watson} W.~D.,  1973, \mn@doi [\apjl] {10.1086/181222}, \href
  {http://adsabs.harvard.edu/abs/1973ApJ...182L..73W} {182, L73}

\bibitem[\protect\citeauthoryear{{Wei{\ss}} et~al.,}{{Wei{\ss}}
  et~al.}{2013}]{weis13}
{Wei{\ss}} A.,  et~al., 2013, \mn@doi [\apj] {10.1088/0004-637X/767/1/88},
  \href {http://adsabs.harvard.edu/abs/2013ApJ...767...88W} {767, 88}

\bibitem[\protect\citeauthoryear{{Wright}, {van Dishoeck}, {Cox}, {Sidher}  \&
  {Kessler}}{{Wright} et~al.}{1999}]{wrig99}
{Wright} C.~M.,  {van Dishoeck} E.~F.,  {Cox} P.,  {Sidher} S.~D.,   {Kessler}
  M.~F.,  1999, \mn@doi [\apjl] {10.1086/311960}, \href
  {http://adsabs.harvard.edu/abs/1999ApJ...515L..29W} {515, L29}

\bibitem[\protect\citeauthoryear{{Yun} \& {Carilli}}{{Yun} \&
  {Carilli}}{2002}]{yun02}
{Yun} M.~S.,  {Carilli} C.~L.,  2002, \mn@doi [\apj] {10.1086/338924}, \href
  {http://adsabs.harvard.edu/abs/2002ApJ...568...88Y} {568, 88}

\bibitem[\protect\citeauthoryear{{Zhang} et~al.,}{{Zhang}
  et~al.}{2018}]{zhan18}
{Zhang} Z.-Y.,  et~al., 2018, \mn@doi [\mnras] {10.1093/mnras/sty2082}, \href
  {https://ui.adsabs.harvard.edu/abs/2018MNRAS.481...59Z} {481, 59}

\bibitem[\protect\citeauthoryear{{da Cunha} et~al.,}{{da Cunha}
  et~al.}{2013}]{dacu13}
{da Cunha} E.,  et~al., 2013, \mn@doi [\apj] {10.1088/0004-637X/766/1/13},
  \href {http://adsabs.harvard.edu/abs/2013ApJ...766...13D} {766, 13}

\bibitem[\protect\citeauthoryear{{van de Voort}, {Quataert},
  {Faucher-Gigu{\`e}re}, {Kere{\v s}}, {Hopkins}, {Chan}, {Feldmann}  \&
  {Hafen}}{{van de Voort} et~al.}{2018}]{vand18}
{van de Voort} F.,  {Quataert} E.,  {Faucher-Gigu{\`e}re} C.-A.,  {Kere{\v s}}
  D.,  {Hopkins} P.~F.,  {Chan} T.~K.,  {Feldmann} R.,   {Hafen} Z.,  2018,
  \mn@doi [\mnras] {10.1093/mnras/sty591}, \href
  {http://adsabs.harvard.edu/abs/2018MNRAS.477...80V} {477, 80}

\makeatother
\end{thebibliography}

\newpage
\appendix
\section{Tentative Detections}\label{TD}

While there are no significant detections of HD(J=1-0)112.2\,$\mu$m line emission in these data, there are four emission features that could be interpreted as tentative detections (see Section \ref{linesec}). Here, we identify the velocity width of each feature, use these velocity ranges to create moment zero maps and integrated spectra, and discuss why they are disregarded as spurious noise peaks. In order to avoid including noise, spectra are extracted from a combination of two spatial masks: $>2\sigma$ in FIR continuum continuum emission and $>3\sigma$ in the moment zero map created using the relevant channels. Velocities are given with respect to the expected redshifted frequency of HD(1-0) at $z=5.656$, or 401.890\,GHz.

\begin{figure}
\centering
\includegraphics[width=\columnwidth]{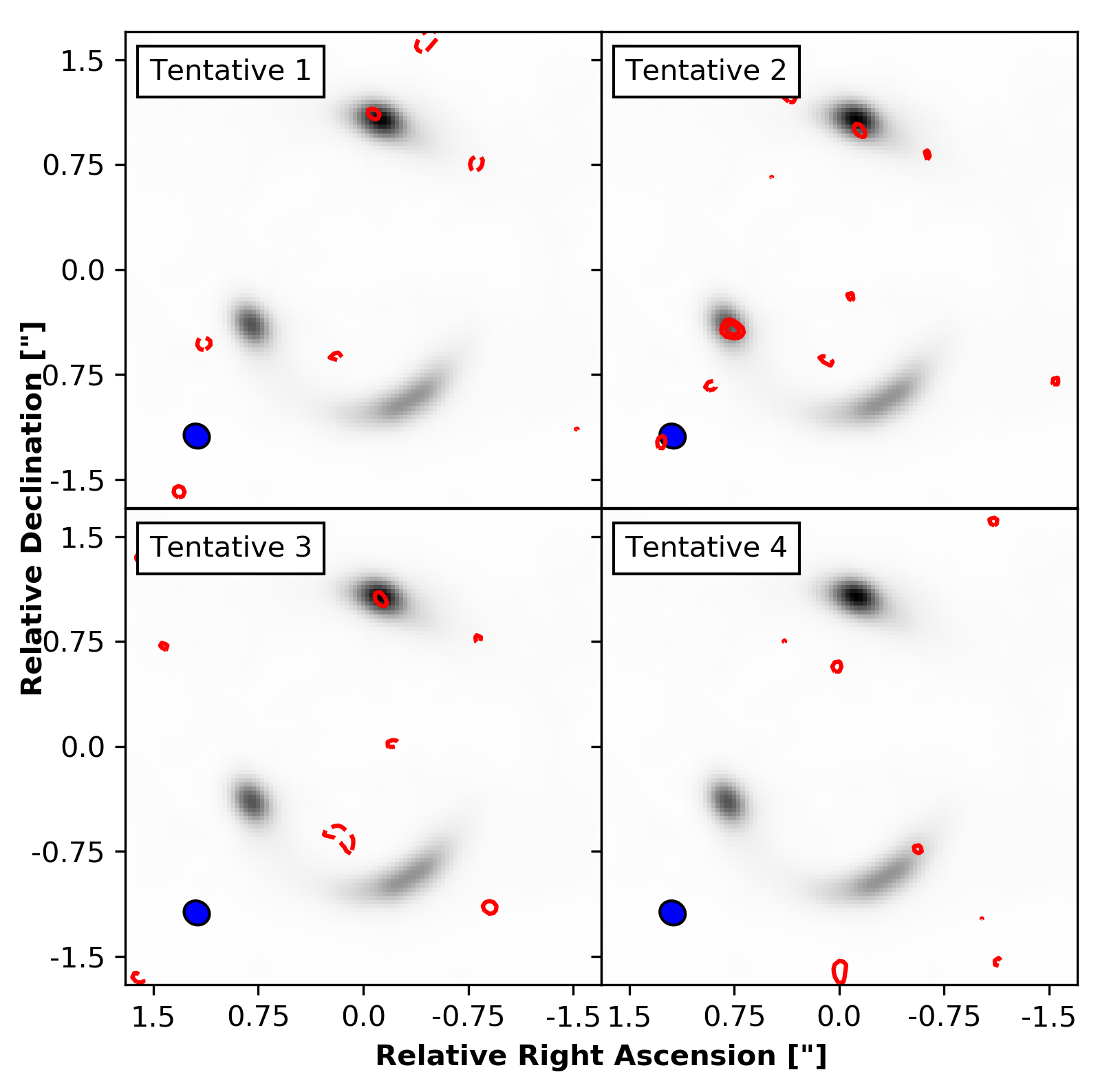}
\caption{Moment zero maps of four tentative line detections (colored contours) compared with FIR continuum (background greyscale). Contours are shown at $-3,3,4\times\sigma$, where $1\sigma=$ 25.0, 18.0, 43.0, and 9.6\,mJy\,beam$^{-1}$\,km\,s$^{-1}$ for Tentative 1 through 4,} respectively.
\label{fig:tentspat1}
\end{figure}

\begin{figure}
\centering
\includegraphics[width=\columnwidth]{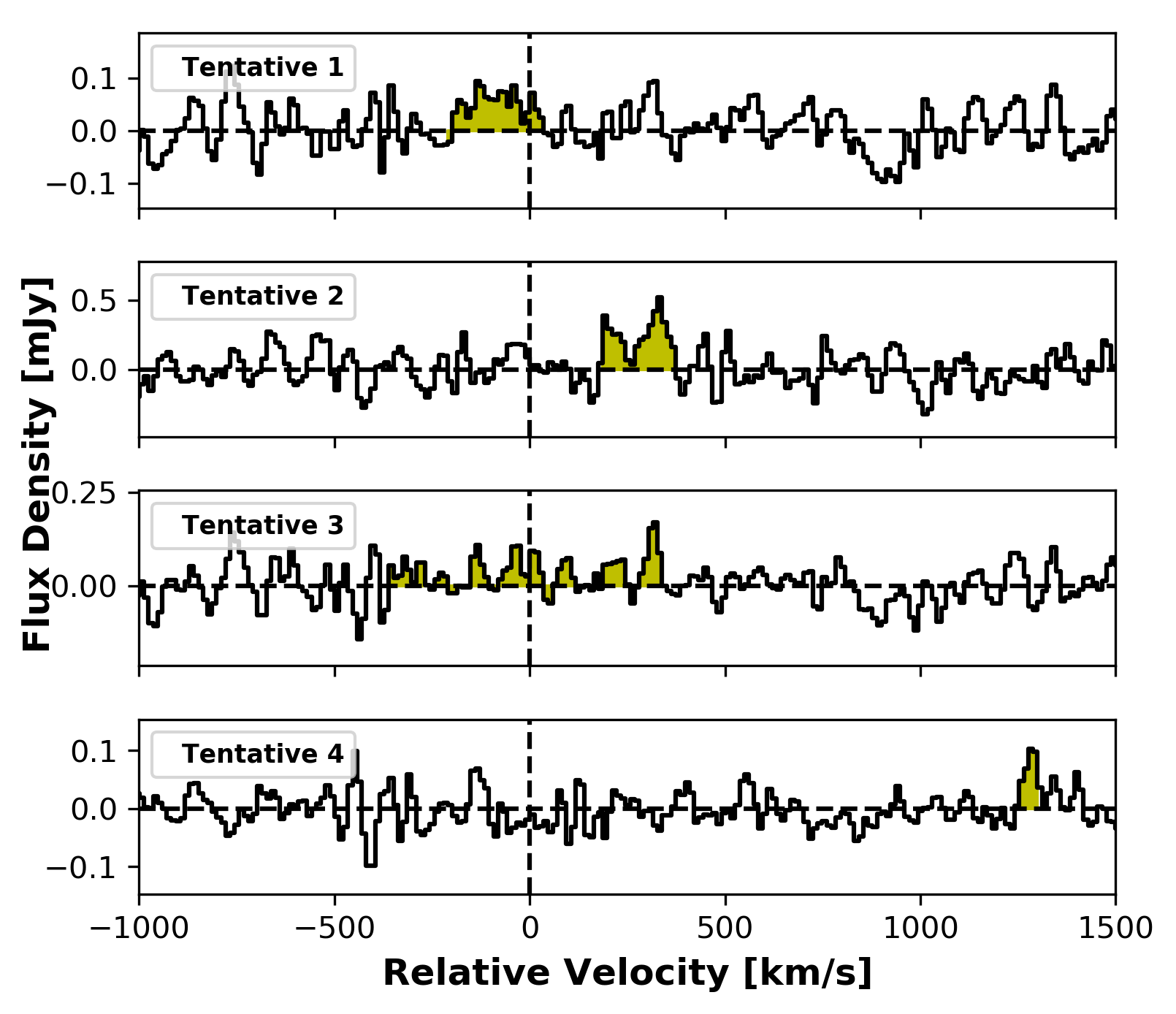}
\caption{Spectra of each tentative line feature in Figure \ref{fig:tentspat1}. Each is extracted using combined masks of $>2\sigma$ in FIR continuum emission and $>3\sigma$ in the moment zero map. The possible line feature is denoted by the yellow shaded region, while the expected redshift of HD(1-0) is shown by a vertical dashed line. No lensing correction has been applied.}
\label{fig:tentspec1}
\end{figure}

The first tentative line feature is detected through an exploration of the continuum-subtracted data cube. It is primarily detected at the northern image ('Tentative 1', -192 to 30\,km\,s$^{-1}$, 402.148 to 401.851\,GHz). The moment zero map of this feature shows $>3\sigma$ emission near the north continuum peak (red contours in top left panel of Figure \ref{fig:tentspat1}), and the emission line itself is broad and has a small velocity offset from \spt (Figure \ref{fig:tentspec1}), so this feature is marginally believable. However, multiple other $3\sigma$ noise peaks are apparent in the collapsed image, so this feature is not confidently detected. 

This exploration of the data cube resulted in an additional feature detected near the southeastern image ('Tentative 2', 193 to 368\,km\,s$^{-1}$, 401.632 to 401.398\,GHz). The collapsed image of these channels shows $>4\sigma$ emission near the southeastern image and $>3\sigma$ emission near the northern image (blue contours of Figure \ref{fig:tentspat1}). However, the spectrum shows a significant velocity offset from \spt ($\sim300$\,km\,s$^{-1}$) and a double-peaked profile, suggesting noise-domination. At first glance, this line shape may be interpreted as evidence for rotation. However, previous studies of line emission have revealed that the FWHM of this source is $\sim600$\,km\,s$^{-1}$ (e.g., \citealt{arav16}), while this feature is $<200$\,km\,s$^{-1}$ wide. Since it has a high velocity offset, $<5\sigma$ significant peaks in the collapsed image, and a non-Gaussian profile, we conclude that this feature is noise.

One possible feature is present in a wide-velocity ($\rm -350<v<350$\,km\,s$^{-1}$) moment zero map ('Tentative 3'). It is exactly coincident with the northern image of the FIR continuum, and is detected at $\sim3\sigma$, suggesting a real detection. However, it is comparable to noise peaks in the field of view, and a spectrum extracted from its $3\sigma$ contour shows a broad velocity range of positive, low-level emission. Since the amplitude of this feature is nearly constant across this range (i.e., no central peak) and no emission is detected from the other lensed images, we believe this increased emission to be an artifact of continuum subtraction due to the very high significance of the northern continuum peak ($\sim185\sigma$; Section \ref{contsec}).

When it is assumed that the spatial distributions of HD(1-0) and FIR continuum emission are equally extended, we find a single spectral feature ('Tentative 4'). This emission is quite narrow (1247 to 1294\,km\,s$^{-1}$, 400.219 to 400.156\,GHz), is close to an atmospheric transmission line, and is at a large velocity offset from the expected redshift of HD(1-0). In addition, the moment zero map shows that there is only one $3\sigma$ peak  present in the southwestern arc. Therefore, this feature is likely only noise.


\label{lastpage}
\end{document}